%% file: main.tex
\documentclass[a4paper,11pt]{article}
\usepackage{pos}
\usepackage{tikz}
\usepackage{cleveref}
\usetikzlibrary{positioning}

\crefname{section}{Sec.}{Secs.}       
\Crefname{section}{Sec.}{Secs.}       
\crefname{figure}{Fig.}{Figs.}
\Crefname{figure}{Fig.}{Figs.}
\crefname{table}{Tab.}{Tabs.}
\Crefname{table}{Tab.}{Tabs.}
\crefname{equation}{Eq.}{Eqs.}

\newcommand{\distro}[4][40]{
  \begin{tikzpicture}[thick]
    \draw[dashed, dash pattern={on 2.3 off 2}] (0, .4) circle (12mm);
    \draw[blue!60!black, very thick] plot[variable=\t, domain=-1:1, samples=#1] ({\t}, {#2 * exp(-10*(\t)^2) + #3 * exp(-60*(\t-0.6)^2 - \t) + #3 * exp(-60*(\t+0.7)^2 - 0.2) + #4 * 0.5 * exp(-50*(\t+0.3)^2) + #4 * exp(-50*(\t-0.2)^2 + 0.1)});
    \draw[solid, ->] (-1, 0)--(1, 0);
    \draw[solid, ->] (0, -0.5)--(0, 1.25);
  \end{tikzpicture}
}

\author*[a,b]{Dominic Schuh}
\author*[a,b]{Janik Kreit}
\author[c,d,e,f]{Evan Berkowitz}
\author[a,b]{Lena Funcke}
\author[b,d]{Thomas Luu}
\author[a,b]{Kim A. Nicoli}
\author[b,d,e,g]{Marcel Rodekamp}

\affiliation[a]{Transdisciplinary Research Area ``Building Blocks of Matter and Fundamental Interactions'' (TRA Matter), University of Bonn, Bonn, Germany}

\affiliation[b]{Helmholtz-Institut f{\"u}r Strahlen- und Kernphysik (HISKP), University of Bonn, Bonn, Germany}

\affiliation[c]{%
J{\"u}lich Supercomputing Center (JSC), Forschungszentrum J{\"u}lich,  J{\"u}lich, Germany
}%

\affiliation[d]{%
Institute for Advanced Simulation 4 (IAS-4), Forschungszentrum J{\"u}lich, J{\"u}lich, Germany
}%

\affiliation[e]{%
Center for Advanced Simulation and Analytics (CASA), Forschungszentrum J{\"u}lich, J{\"u}lich, Germany
}%

\affiliation[f]{
   College of Science and Mathematics, University of the Virgin Islands, RR1 Box 10000 Kingshill, St. Croix 00850, US Virgin Islands
}%

\affiliation[g]{Institut für Theoretische Physik, Universit{\"a}t Regensburg, 93053 Regensburg, Germany}

\emailAdd{schuh@hiskp.uni-bonn.de}
\emailAdd{kreit@hiskp.uni-bonn.de}
\emailAdd{evan.berkowitz@uvi.edu}
\emailAdd{lfuncke@uni-bonn.de}
\emailAdd{t.luu@fz-juelich.de}
\emailAdd{knicoli@uni-bonn.de}
\emailAdd{marcel.rodekamp@ur.de}

\title{Simulating the Hubbard Model with Equivariant Normalizing Flows}

\abstract{Generative models, particularly  normalizing flows, have shown exceptional  performance in learning probability distributions across various domains of physics, including statistical mechanics, collider physics, and lattice field theory. In the context of lattice field theory, normalizing flows have been successfully applied to accurately learn the Boltzmann distribution, enabling a range of tasks such as direct estimation of thermodynamic observables and sampling independent and identically distributed (i.i.d.) configurations.\newline
In this work, we present a proof-of-concept demonstration that normalizing flows can be used to learn the Boltzmann distribution for the Hubbard model. This model is widely employed to study the electronic structure of graphene and other carbon nanomaterials. State-of-the-art numerical simulations of the Hubbard model, such as those based on Hybrid Monte Carlo (HMC) methods, often suffer from ergodicity issues, potentially leading to biased estimates of physical observables. Our numerical experiments demonstrate that leveraging i.i.d.\ sampling from the normalizing flow effectively addresses these issues.}

\FullConference{%
  The 41st International Symposium on Lattice Field Theory (Lattice2024)\\
  29 July-02 August 2024\\
  Liverpool, United Kingdom
}

\begin{document}
\maketitle

\section{Introduction}
\noindent
The Hubbard model is a condensed matter model describing the interaction of strongly coupled electronic systems. Studying this model often poses severe challenges for numerical simulations, primarily due to its multi-modal nature and intricate energy landscape. Standard sampling methods, such as Hybrid Monte Carlo (HMC), often suffer from ergodicity problems~\cite{PhysRevB.100.075141}, potentially leading to biased and inaccurate estimates of physical observables. Novel techniques have been developed towards having more efficient samplers, such as the combination of HMC with radial updates \cite{temmen2024overcomingergodicityproblemshybrid}.

In these proceedings, we deploy a deep generative machine learning approach to overcome the ergodicity issue in the context of the Hubbard model. Deep generative models, also known as generative neural samplers (GNSs), in particular normalizing flows~\cite{rezende2015variational,kobyzev2020normalizing,nfreview} and autoregressive neural networks~\cite{oord2016pixelrecurrentneuralnetworks,NIPS2016_b1301141}, have demonstrated great capabilities for modeling Boltzmann distributions of physical and chemical systems. These generative approaches, often known as Boltzmann generators~\cite{doi:10.1126/science.aaw1147}, have lately emerged in various fields, ranging from lattice quantum field theory~\cite{PhysRevD.100.034515,PhysRevLett.126.032001,Caselle:2022acb,cranmer2023advances} to statistical mechanics~\cite{PhysRevLett.122.080602,PhysRevE.101.023304} to string theory~\cite{Caselle:2023uel,Caselle:2023mvh} to quantum chemistry~\cite{doi:10.1126/science.aaw1147,gebauer1,gebauer3}. Generative neural samplers hold promise not only for allowing faster and more efficient sampling, but also for directly estimating thermodynamic observables~\cite{PhysRevLett.126.032001} as well as entanglement entropies~\cite{bialas2024r,Bulgarelli:2024yrz}.  
Furthermore, recent studies demonstrate that GNSs are highly effective in sampling from distributions with challenging topologies, such as bimodal distributions~\citep{Nicoli:2021inv,PhysRevD.108.114501} and gauge theories affected by topological freezing~\cite{PhysRevLett.125.121601}. This positions GNSs as a well-suited alternative for sampling from challenging probability densities where HMC faces ergodicity issues.

Our work represents the first attempt in applying GNSs to the Hubbard model. Specifically, we propose to use an equivariant normalizing flow~\cite{kohler2020equivariant} to efficiently learn the underlying probability distribution. This approach allows us to directly incorporate an arbitrary number of symmetries in the neural network design.

The remainder of this paper is organized as follows: First, in \cref{sec:hubbard}, we introduce the Hubbard model and discuss how standard HMC methods struggle with ergodicity. In~\cref{sec:flow}, we introduce equivariant normalizing flows and elaborate on their application to the Hubbard model. We present numerical experiments in~\cref{sec:results} to prove the advantages of the proposed method compared to standard HMC. Finally, we provide conclusions and an outlook in~\cref{sec:conclusion}.
\section{Hubbard Model} \label{sec:hubbard}
\subsection{Theoretical Background}
\noindent

The Hubbard model~\cite{hubbard} describes the interaction of fermions on a spatial lattice of fixed ions, where the atomic grid is considered static. It is capable of describing a vast amount of physical systems~\cite{Arovas_2022}, most widely applied to model carbon nanostructures and other graphene formations. 

The dynamics of the Hubbard model at half-filling, i.e., with vanishing chemical potential, are described by the Hamiltonian
\begin{align}
    H_\alpha &= H_{\rm TB} + \alpha \frac{U}{2} \sum_x \left(n_{x,\uparrow} - n_{x, \downarrow} \right) - (1-\alpha) \frac{U}{2} \sum_x \left(n_{x,\uparrow} - n_{x,\downarrow} \right)^2, \label{eq:hubb_ham}\\
    H_{\rm TB} &= -\kappa \sum_{\langle x,y \rangle} \left( a_x^\dagger a_y - b_x^\dagger b_y \right), \label{eq:TB_ham}
\end{align}
where~\Cref{eq:TB_ham} is often referred to as the \textit{tight-binding} Hamiltonian. It describes the binding and hopping of particles between adjacent atomic sites in a solid state, with $\kappa$ being responsible for the hopping dynamics. The operators $a_x^\dagger (a_x)$ and $b_x^\dagger (b_x)$ create (annihilate) a spin-up and spin-down particle, respectively, with the  number operator $n_{x,\uparrow}=a_x^\dagger a_x$ for a spin-up particle. In~\Cref{eq:hubb_ham}, the parameter $U$ describes the on-site interaction strength, while $\alpha \in \left[0,1\right]$ continuously parametrizes the choice of basis, i.e., $\alpha = 0$ denotes the spin basis and $\alpha = 1$ the particle/hole basis. Thus, the parameter $\alpha$ defines the physical interpretation of the degrees of freedom, i.e., particles with spin up or down versus particles and holes, respectively. In this work, we will restrict ourselves to the particle/hole (ph) basis, while in future work we will extend our analysis to the spin basis.

A probabilistic interpretation of the partition function 
\begin{align}
    Z_{\rm ph} = \mathrm{tr} \left(e^{-\beta H_{\alpha=1}}\right)
\end{align}
requires the transition to a path integral formulation, which is achieved by discretizing the inverse temperature $\beta$ into $N_t$ evenly spaced, so-called \textit{time-slices} and performing a Suzuki-Trotter decomposition of second order \cite{trotter1959product, 10.1143/PTP.56.1454}. Furthermore, to decouple a quartic fermionic term appearing in the path integral, a continuous Hubbard-Stratonovich (HS) transformation \cite{Hubbard:1959ub} is needed,
\begin{align}\label{eq:HS}
    e^{-\frac{1}{2}\tilde{U}n^2} = \frac{1}{\sqrt{2\pi \tilde{U}}} \int_{-\infty}^{\infty}\, \mathrm{d}\phi \,e^{-\frac{1}{2\tilde{U}}\phi^2 \pm i\phi n},
\end{align}
where $\tilde{U}\equiv U\beta/N_t$ is the rescaled on-site interaction strength, $n=n_{\downarrow}-n_{\uparrow}$ is a fermionic number operator, and $\phi$ is a bosonic auxiliary field. This transformation reduces the power of the Grassmanian fields from four to two at the cost of introducing one bosonic auxiliary field per fermionic degree of freedom. Lastly, by inserting coherent fermionic states and integrating out the remaining fermionic degrees of freedom using standard Gaussian integrals, the partition function is rendered purely dependent on the bosonic fields \cite{brower2012hybridmontecarlosimulation, Ulybyshev_2013, Smith_2014, Luu_2016},
\begin{align}\label{eq:partition}
    Z_{\rm ph}  = \int_{-\infty}^{\infty} \left[ \prod_{x,t} \mathrm{d}\phi_{xt}\right]e^{-S[\phi]},
\end{align}
with the Hubbard action  defined as
\begin{align}\label{eq:HM_action}
    S[\phi] = \frac{1}{2\Tilde{U}}\sum_{x,t\in\Lambda} \phi_{xt}^2 - \log \det M \left[i\phi \right] - \log \det M [-i\phi],
\end{align}
where $\Lambda=N_x\times N_t$ is the space-time volume of the lattice and $xt$ represents a site on the lattice.
The fermion matrix $M$ in the exponential discretization\footnote{There are other choices for incorporating the hopping term in the fermion matrix, such as the diagonal \cite{brower2012hybridmontecarlosimulation, Luu_2016} or the linear \cite{Smith_2014} discretization. However, all discretizations agree in the continuum limit $N_t\rightarrow \infty$ \cite{PhysRevB.100.075141}.} reads
\begin{align}\label{eq:fermion_matrix}
    M^e \left[ \phi \right]_{x't',xt} = \delta_{x',x}\delta_{t',t} - [e^h]_{x',x} e^{\phi_{xt}} \mathcal{B}_{t'} \delta_{t',t+1},
\end{align}
with the hopping matrix $h_{x',x}=\kappa \delta_{\langle x',x\rangle}$. Here, $\mathcal{B}$ explicitly incorporates anti-periodic boundary conditions in the temporal direction $t$, i.e., $\mathcal{B}_{t} = +1$ for $0<t<N_t$ and $\mathcal{B}_0=-1$. Taking a closer look at the Hubbard action in~\Cref{eq:HM_action}, we see that it contains two different components: a Gaussian term and a fermion determinant. The first one stems from the Hubbard-Stratonovich transformation in~\cref{eq:HS} and accounts for the on-site interaction previously introduced in~\Cref{eq:hubb_ham}, while the latter one encodes the hopping dynamics described by the tight-binding Hamiltonian in~\Cref{eq:TB_ham}.

In these proceedings, we focus on a $1+1$D lattice, where the spatial lattice extent is fixed to two, i.e. $N_x=2$, and we consider different temporal lattice extents, $N_t$. Unless specified otherwise, we work in the particle/hole basis with the fermion matrix in the exponential discretization.

\subsection{Symmetries of the Hubbard Action}\label{sec:symm}
\noindent
The Hubbard action in~\cref{eq:HM_action} obeys a large set of symmetries. Generally, these can be symmetries of the entire action or symmetries only preserving the fermion determinant. For a comprehensive discussion of the symmetries of the Hubbard model, we refer the reader to~\cite{PhysRevB.100.075141}. In the following, we outline the specific symmetries that will be relevant for these proceedings:

\begin{itemize}\item{\textbf{$\mathbf{Z}_2$-Symmetry}}: The entire action is invariant under a sign-flip transformation
    \begin{equation*}
    \phi \rightarrow -\phi,
\end{equation*}
i.e., in this choice of basis, the action is invariant under the exchange of particles and holes. Note that this is only a symmetry on bipartite lattices at half-filling.

\item{\textbf{Space-Translation Symmetry}}:
Under the condition that the on-site interaction strength $U$ is uniform across the entire lattice, the action is invariant under the exchange of \textbf{all} spatial sites, i.e.,
\begin{equation*}
    (\phi_{x_1}, \phi_{x_2}) \rightarrow (\phi_{x_2}, \phi_{x_1}).
\end{equation*}
\item{\textbf{Periodicity Symmetry}}:
The only symmetry of relevance that is not preserved by the Gaussian part of the action is the $2\pi$ periodicity symmetry of the fermion determinant \cite{PhysRevB.100.075141}. As can be seen in \Cref{eq:fermion_matrix,eq:HM_action}, the determinant $M^e[i\phi]$ only depends on $e^{i\phi}$ and is therefore invariant under
\begin{align*}
    \phi_{xt} &\rightarrow \phi_{xt} + n\cdot 2\pi, & n&\in \mathcal{Z}.
\end{align*}
\noindent
\end{itemize}
These symmetries will become important in~\Cref{sec:Eq_NF} for designing equivariant normalizing flows.

\subsection{Hybrid Monte Carlo Simulations}
\begin{figure}[t!]
    \centering
    \includegraphics[width=0.49\linewidth]{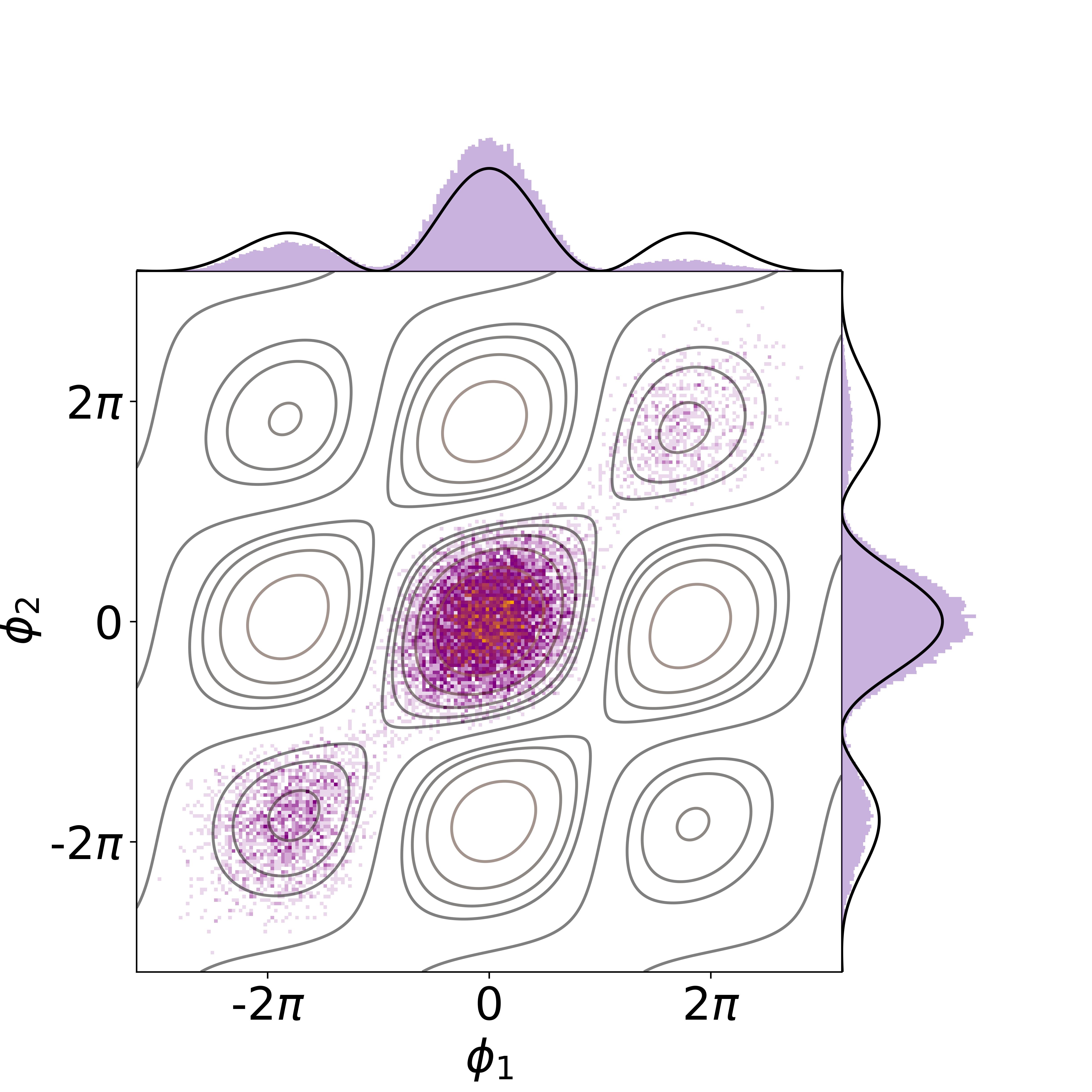}
    \includegraphics[width=0.49\linewidth]{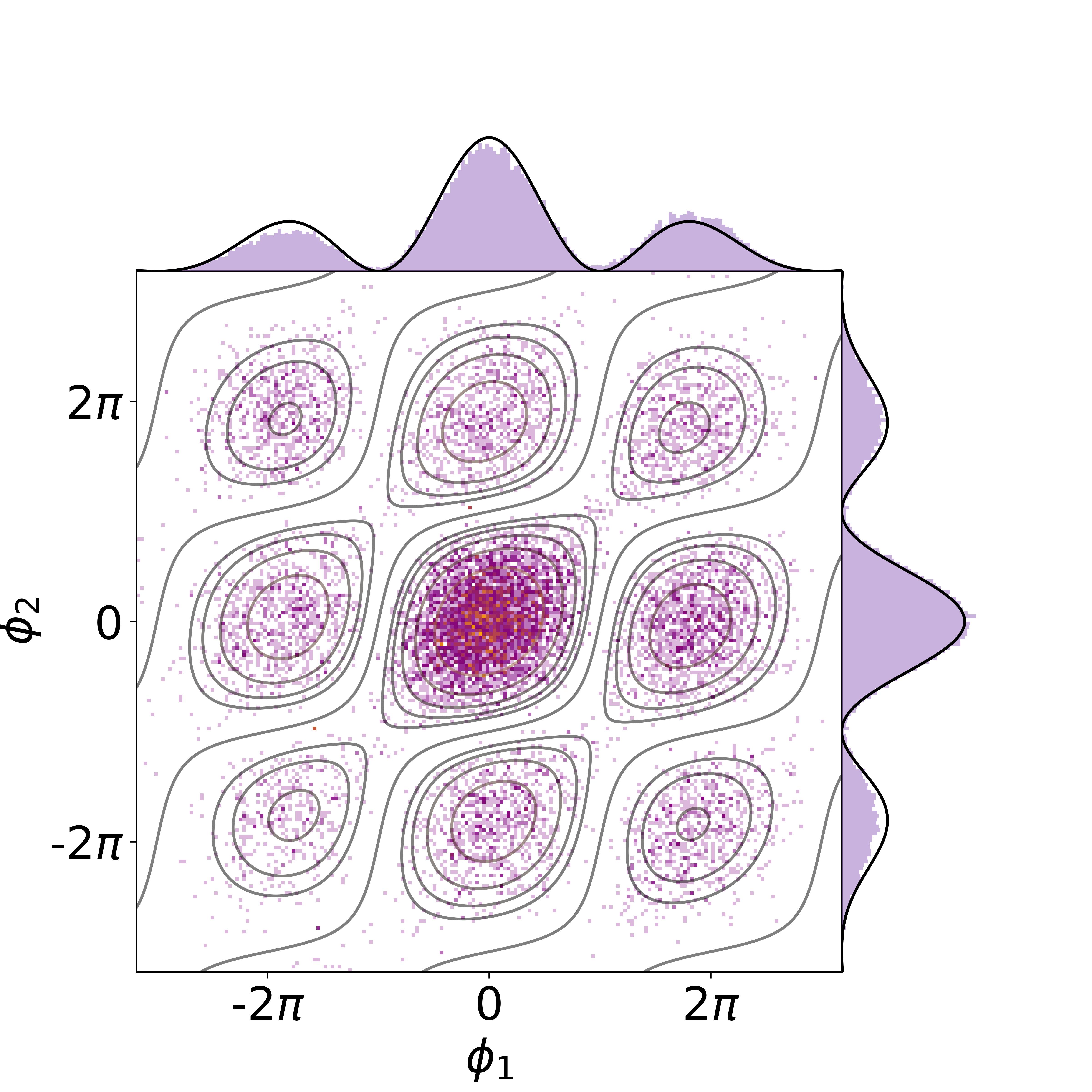}
    \caption{HMC configurations for a $(N_x,N_t) = (2,1)$ lattice with parameters $\beta=\kappa=1$ and $U=18$, generated with a leapfrog integrator with an integration step of $\epsilon = 0.1$ (left) and $\epsilon = 1.0$ (right). The integrated autocorrelation time $\tau_\mathrm{int}$ and the acceptance rate $a$ are $\tau_{\mathrm{int}} = 83 \pm 14$ and $a = 99.9\%$ (left) and  $\tau_\mathrm{int}= 346\pm100$ and $a = 86.9\%$ (right), respectively. Histograms of the magnetization, obtained by summing along each dimension, respectively, are shown at the top and on the right hand side of the figures. The analytical contours are exact for $N_t=1$ and in the strong-coupling limit, i.e. $U/\kappa\rightarrow \infty$,  for $N_t>1$. The colormap ranging from purple to yellow represents the areas of low and high density, respectively.} 
    \label{fig:hmc}
\end{figure}

\noindent
One widely used method to generate samples for the Hubbard model is the Hybrid Monte Carlo (HMC) algorithm~\cite{Duane:1987de}. While HMC is extremely successful in various applications, it faces substantial challenges when being applied to the Hubbard model. As illustrated in~\Cref{fig:hmc}, the density for a $(N_x, N_t) = (2,1)$ lattice consists of diagonal bands separated by infinite potential barriers, with an underlying periodic blob structure, ``wrapped'' by a Gaussian. The multi-modal nature of this distribution presents significant ergodicity challenges, as the leapfrog integrator is repelled by high potential barriers. This makes it hard to tunnel between the modes and reach every point in configuration space, as shown in the left panel of~\Cref{fig:hmc}. Furthermore, the separation of the different regions in the $N_xN_t$-dimensional configuration space is determined by manifolds where the fermion determinant vanishes. These manifolds are of codimension 1 \cite{PhysRevB.100.075141}, i.e., they are of dimension $N_xN_t-1$, and, as a result, are present at all lattice sizes.

This issue may be addressed by a coarser integrator since the proposed configurations in the Markov chain have a larger spread. However, a coarser integrator leads to a substantial decrease in the acceptance rate, along with an increase in the integrated autocorrelation time $\tau_\mathrm{int}$, as demonstrated in the right panel of~\Cref{fig:hmc}.  

\section{Normalizing Flows}\label{sec:flow}
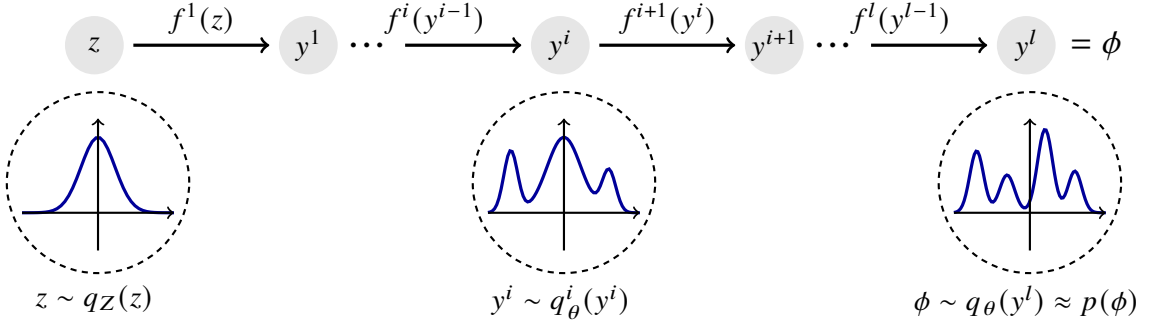
\begin{figure}[tb]
    \centering
    \input{flow}
    \caption{Schematic representation of a coupling-based normalizing flow. Starting with a Gaussian prior distribution $q_Z(z)$ (left), the normalizing flow $f_\theta(z)$ maps the prior samples $z$ to the target samples $\phi$ through a set of invertible functions $f^i$, each taking as input the output of the previous layer, $y^{i-1}$, ultimately approximating the target distribution $p(\phi)$ (right). In the context of lattice field theories, this target distribution corresponds to the path integral distribution $p(\phi) = Z^{-1} e^{-S[\phi]}$, where $S[\phi]$ is the action of the theory. 
    }
    \label{fig:flow}
\end{figure}

\noindent
A normalizing flow~\cite{kobyzev2020normalizing,nfreview} is a parametric bijective map, $f_\theta: \mathcal{Z}, \rightarrow \mathcal{K}$ from a latent space $\mathcal{Z}$ with a simple prior distribution $q_Z(z)$, e.g., a Gaussian, to a target space $\mathcal{K}$ with a target distribution $p(\phi)$. In the case of coupling-based normalizing flow~\cite{dinh2014nice,dinh2017densityestimationusingreal}, this map consists of a series of invertible and differentiable transformations $f^i$ (see~\Cref{fig:flow}), each parametrized by neural networks with parameters $\theta_i$. This ensures bijectivity of the entire map, yielding
\begin{align}
    f_\theta(z) = \left( f^l \circ f^{l-1} \circ \dots \circ f^1 \right) (z)\,,
\end{align}
where the dependence on the parameters $\theta_i$ for each block was omitted for ease of notation. 
During the training of a normalizing flow, the parameters $\theta_i$ of the neural networks parametrizing each block $f^i_{\theta_i}$ are optimized with the goal of enabling the normalizing flow to approximate a target distribution $p(\phi)$. Once the flow has been trained, sampling becomes efficient, as it only requires drawing samples from the base distribution $q_Z(z)$ and transforming them according to the parametrized map $f_\theta(z)$, thus allowing for embarrassingly parallelizable generation of independent and identically distributed (i.i.d) samples from the approximated density $q_\theta(z)\approx p(\phi)$.

When considering lattice field theories, the goal is to generate samples from the path integral distribution of the theory. Therefore, the target density to sample from is given by $p(\phi) = Z^{-1}e^{-S[\phi]}$, where $Z$ is the partition function. To approximate this distribution, the parameters $\theta$ of the normalizing flow are optimized by gradient descent while minimizing the so-called reverse Kullback-Leibler divergence~\cite{10.1214/aoms/1177729694}

\begin{align}\label{eq:KL}
    \mathrm{KL}(q_\theta ||p) = \int \mathcal
D[\phi] q_\theta(\phi) \ln\left(\frac
{q_\theta(\phi)}{p(\phi)}\right),
\end{align}
which provides a measure of how much the variational distribution $q_\theta$ generated by the normalizing flow differs from the target distribution $p$. The likelihood of $q_\theta$ is known analytically,
\begin{align}
    q_{{\theta}}(\phi) &= q_z(f^{-1}_{{\theta}}(\phi)) \, \left| \frac{\textrm{d}f_{{\theta}}}{\textrm{d}{z}} \right|^{-1} = q_z({z}) \, \left| \frac{\textrm{d}f_{{\theta}}}{\textrm{d}{z}} \right|^{-1}\,,
\end{align}
which allows to rewrite~\cref{eq:KL} as
\begin{align}
    \mathrm{KL}(q_\theta ||p) = \mathbb{E}_{z\sim q_Z} \left[S[f_\theta(z)]- \ln \left| \frac{
\mathrm{d}f_\theta}{\mathrm{d}z}\right|(z) + \ln q_Z(z) +\ln Z\right].
\end{align}
This expression contains the logarithm of the partition function, which is unknown. However, being a constant shift, this quantity vanishes upon taking the gradient with respect to the model parameters. 

Each component of $f_\theta$ has to be designed such that each transformation is invertible and has a tractable Jacobian to ensure the efficient computation of its determinant. Many architectures have been proposed that fulfill these requirements, we refer to Ref.~\cite{kobyzev2020normalizing} for an overview. In this work, we use the so-called \textit{real-valued non-volume preserving} (RealNVP)~\cite{dinh2017densityestimationusingreal} transformation. This particular type of affine transformation requires two neural networks $s_\theta$ and $t_\theta$, responsible for scaling and shifting the input $y^l$, respectively. Furthermore, to preserve the invertibility of the transformation, the input sites $y^l$ are divided into two partitionings, $y_\mathrm{on}^l$ and $y_\mathrm{off}^l$, and the transformation acting on the \textit{on} sites is restricted to depend on the complementary \textit{off} sites,
\begin{align}
\begin{aligned}
    y^{l+1}_\mathrm{off} &= y^l_\mathrm{off}, \\
    y^{l+1}_\mathrm{on} &= y^l_\mathrm{on} \cdot \exp\left[s_\theta\left(y^l_\mathrm{off} \right) \right] + t_\theta\left(y^l_\mathrm{off} \right)\,. 
\end{aligned}
\end{align}
This type of transformation ensures a tractable Jacobian determinant
\begin{align}
    \det \frac{\partial y^{l+1}}{\partial y^l} &= 
    \begin{vmatrix}
        \mathbf{I}_d & 0 \\
        * & \text{diag} \left( \exp \left[ s_\theta \left( y^l_\mathrm{off} \right) \right] \right)
    \end{vmatrix}\,,
\end{align}
where $d$ refers to the dimensionality of the \textit{off} partitioning and $y^l$ is the output of the $l$-th layer.

Every block $f^i(y^{i-1})$, as shown in~\cref{fig:flow}, is an instance of a RealNVP transformation, i.e., a RealNVP block, with an alternating splitting between \textit{on} and \textit{off} partitionings, to ensure that the entire input is transformed within two blocks.

\subsection{Equivariant Normalizing Flows for the Hubbard Model}\label{sec:Eq_NF}
\noindent
Equivariant normalizing flows~\cite{kohler2020equivariant, boyda2021sampling} have been proposed to integrate prior knowledge, e.g., known symmetries, of the target distribution into the deep generative model, in order to make the bijective map equivariant by design. Reducing redundancies in the distribution in this way leads to more efficient training as well as more accurate results.

Let $T$ be some transformation that represents a symmetry of the target distribution. A neural network $f_\theta(z)$ is \textit{equivariant} with respect to $T$ if it fulfills~\cite{kohler2020equivariant}
\begin{align}
    f_\theta(z) = T^{-1}f_\theta(Tz).
\end{align}
This way, the input $z$ gets transformed into the canonical cell of the theory, leaving the neural network with the task of learning only the canonicalized distribution.

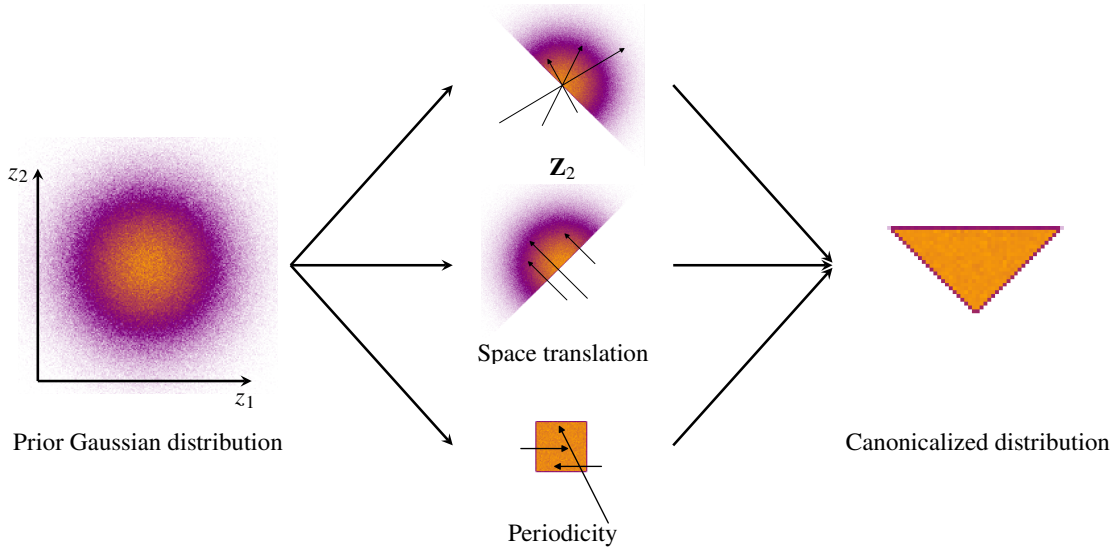
\begin{figure}[tb]
    \centering
    \resizebox{0.99\textwidth}{!}{
    \input{symmetry_transformation}
    }
    \caption{Transformation of the prior distribution into the canonical cell of the theory. Starting from a Gaussian prior distribution (left), we apply three types of symmetry transformations (middle): $\mathbf{Z}_2$ symmetry (top), space-translation symmetry (center), and periodicity symmetry (bottom). This results  in a canonical cell of triangular shape (right). The small black arrows indicate the mapping of three exemplary samples under these transformation, indicating their position before and after the transformation has been applied. }
    \label{fig:equivariant}
\end{figure}

In the case of the Hubbard model, we leverage the symmetries introduced in Sec.~\ref{sec:symm} and define the following set of transformations:
\begin{align}\mathbf{Z}_2\text{-Symmetry}&\quad : \quad & (z_1,z_2) & \mapsto 
        \begin{cases}
        (z_1,z_2) & \text{if } z_1 + z_2 \geq 0 \\
        -(z_1,z_2) & \text{else}
        \end{cases}
    \\
    \text{Space-Translation~Symmetry}&\quad : \quad &(z_1, z_2) & \mapsto 
        \begin{cases}
        (z_1, z_2) & \text{if } z_1 - z_2 \leq 0 \\
        (z_2, z_1) & \text{else}
        \end{cases} 
    \\
    \mathrm{Periodicity~Symmetry}&\quad : \quad &
    z_{xt} &\mapsto z_{xt} - 2\pi \cdot k, 
    \quad k = \text{round} \left( \frac{z_{xt}}{2\pi} \right),
\end{align}
where $z$ is a sample taken from the Gaussian distribution and $z_x = \sum_{t=1}^{N_t}z_{xt}$. An intuitive understanding of how these symmetries act on the base distribution is illustrated in~\Cref{fig:equivariant}: The periodicity symmetry maps every sample into a square of length $2\pi$, while the $\mathbf{Z_2}-$ and space-translation symmetries \textit{fold} this square, resulting in a triangle. Inverting these symmetries can therefore be understood as \textit{unfolding} this triangle and mapping every sample back to its origin, i.e., going backward by applying the inverse transformation $T^{-1}$.

\section{Results}\label{sec:results}
\noindent
In this section, we apply the aforementioned techniques to the Hubbard model by training a normalizing flow $f_\theta$ to approximate the target distribution $p(\phi)=Z^{-1}e^{-S[\phi]}$, where $Z$ and $S[\phi]$ are given by~\Cref{eq:partition} and~\Cref{eq:HM_action}, respectively. As there is no guarantee that $q_\theta \equiv p$, one should enforce asymptotic unbiasedness~\cite{PhysRevE.101.023304}. To this end, one could either use the so-called NeuralMCMC or metropolization technique by applying an accept-reject step to the i.i.d.\ samples from the normalizing flow, or perform neural importance sampling (NIS)~\cite{PhysRevLett.126.032001}. The numerical experiments shown in these proceedings are based on NeuralMCMC and have been obtained using a preliminary version of the NeuLat software framework~\cite{Nicoli:2023rcd}.

\noindent
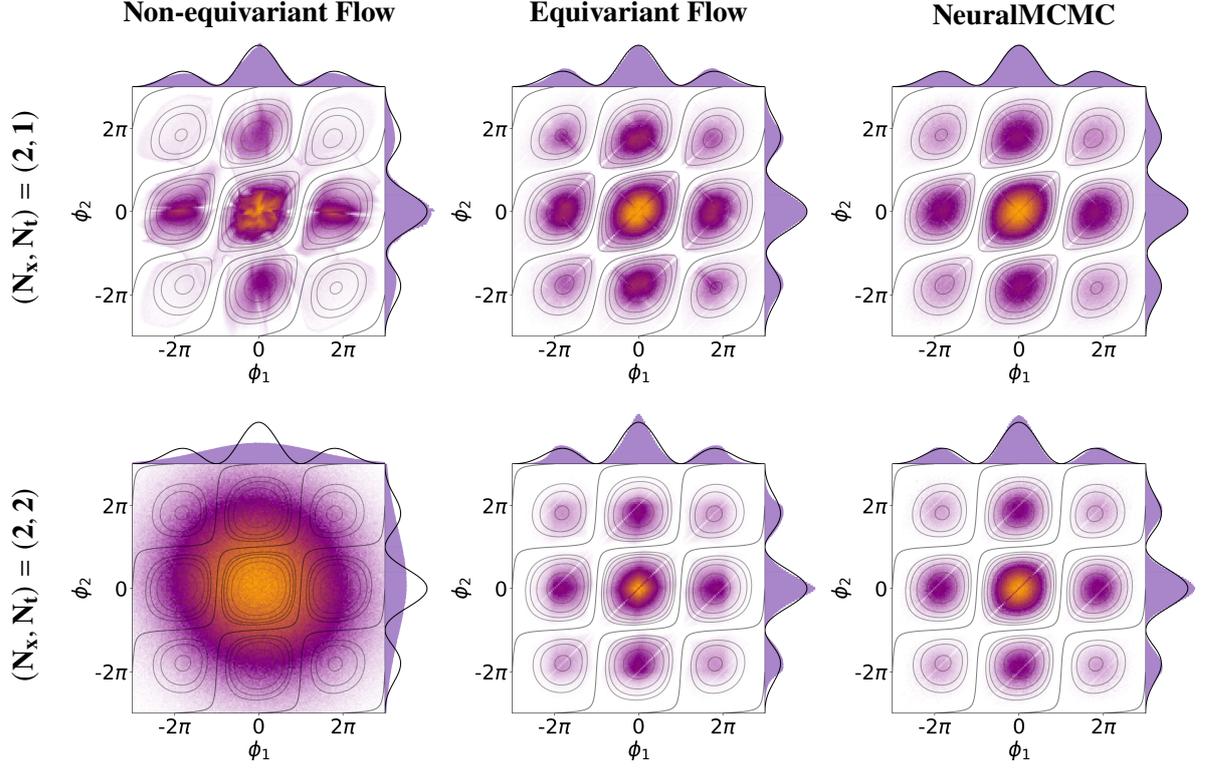
\begin{figure}
\begin{tikzpicture}
\node at (3, 4.5) {\includegraphics[width=0.325\linewidth]{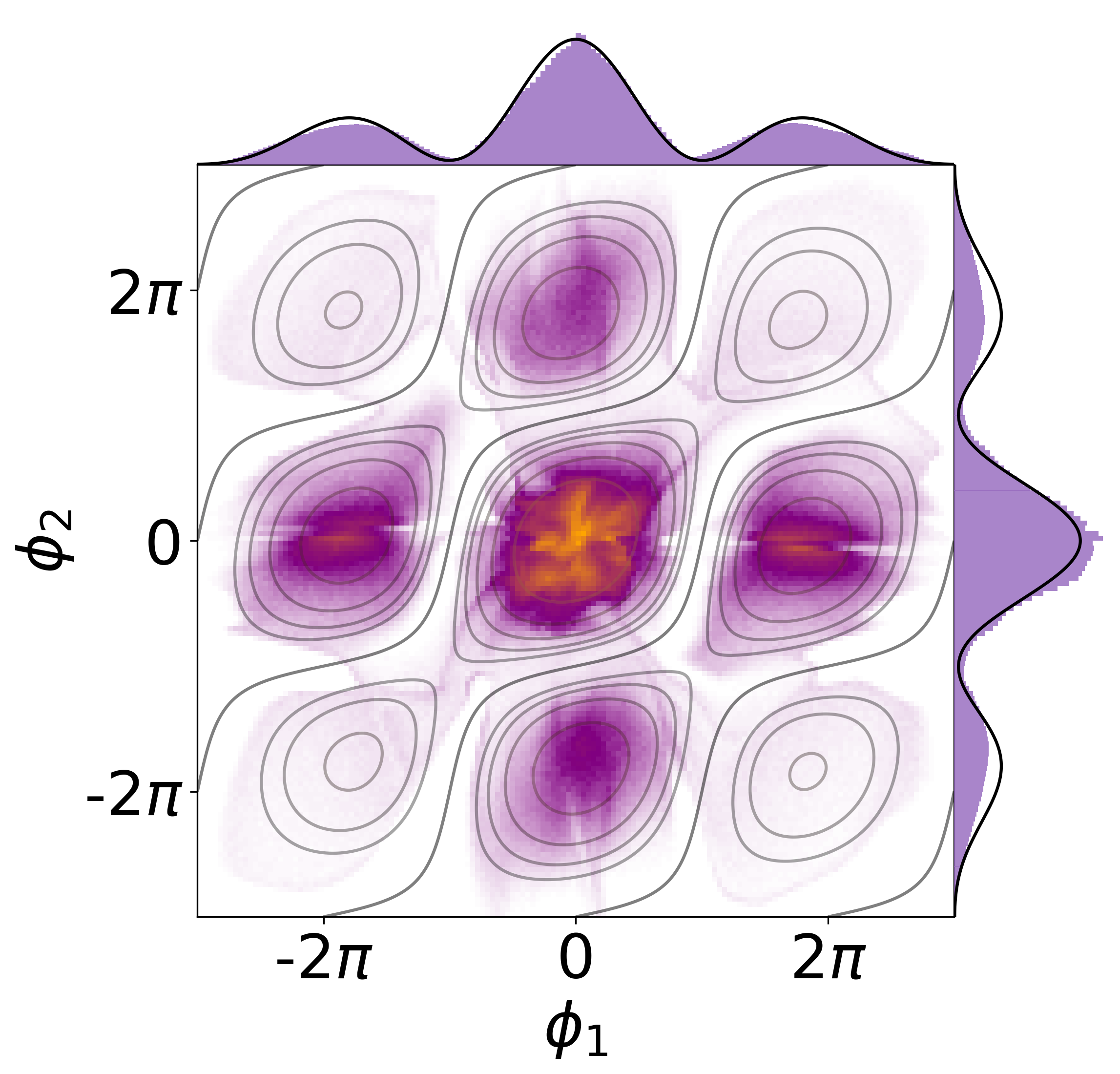}}; 
\node at (8, 4.5) {\includegraphics[width=0.325\linewidth]{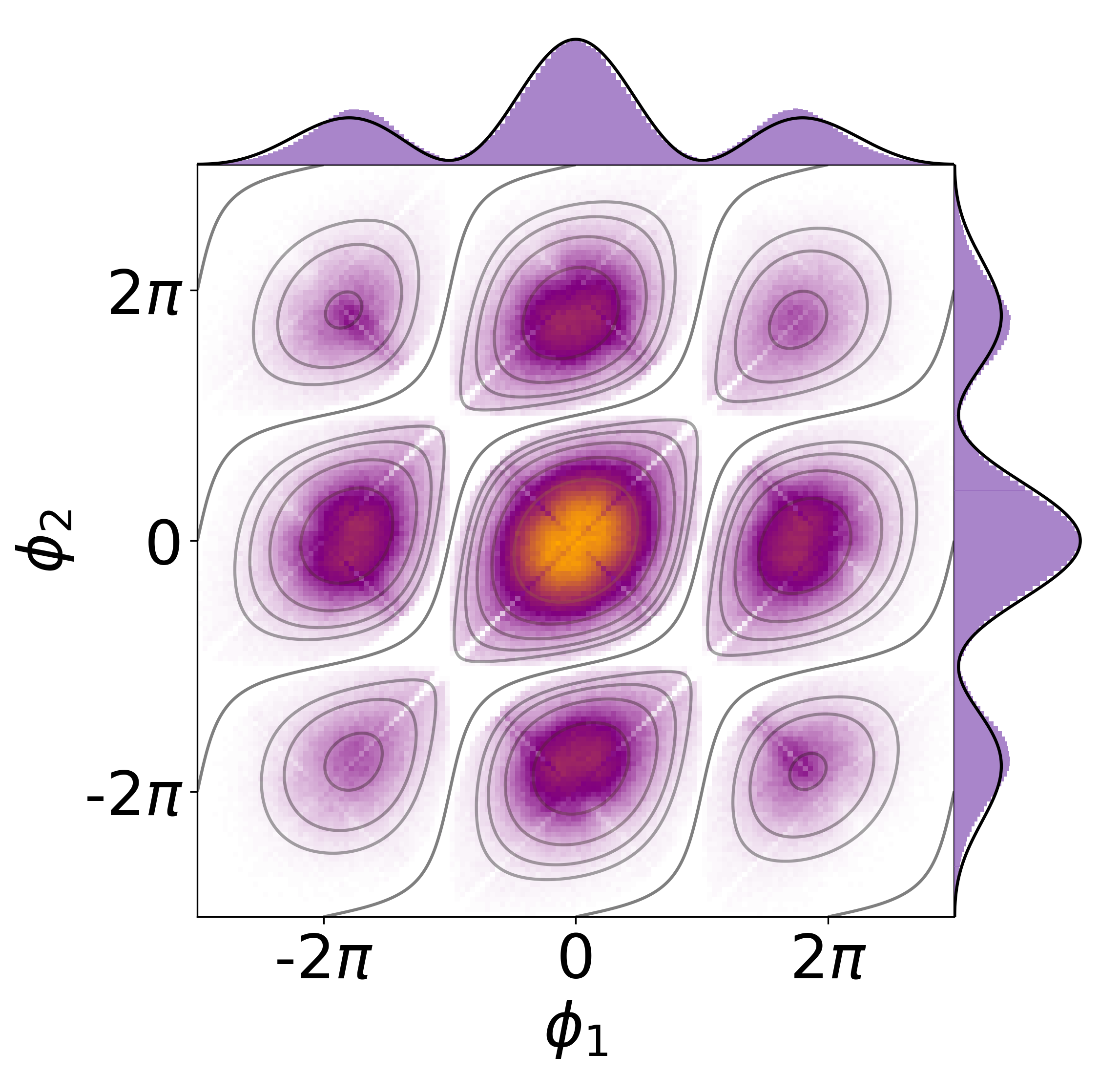}}; 
\node at (13, 4.5) {\includegraphics[width=0.325\linewidth]{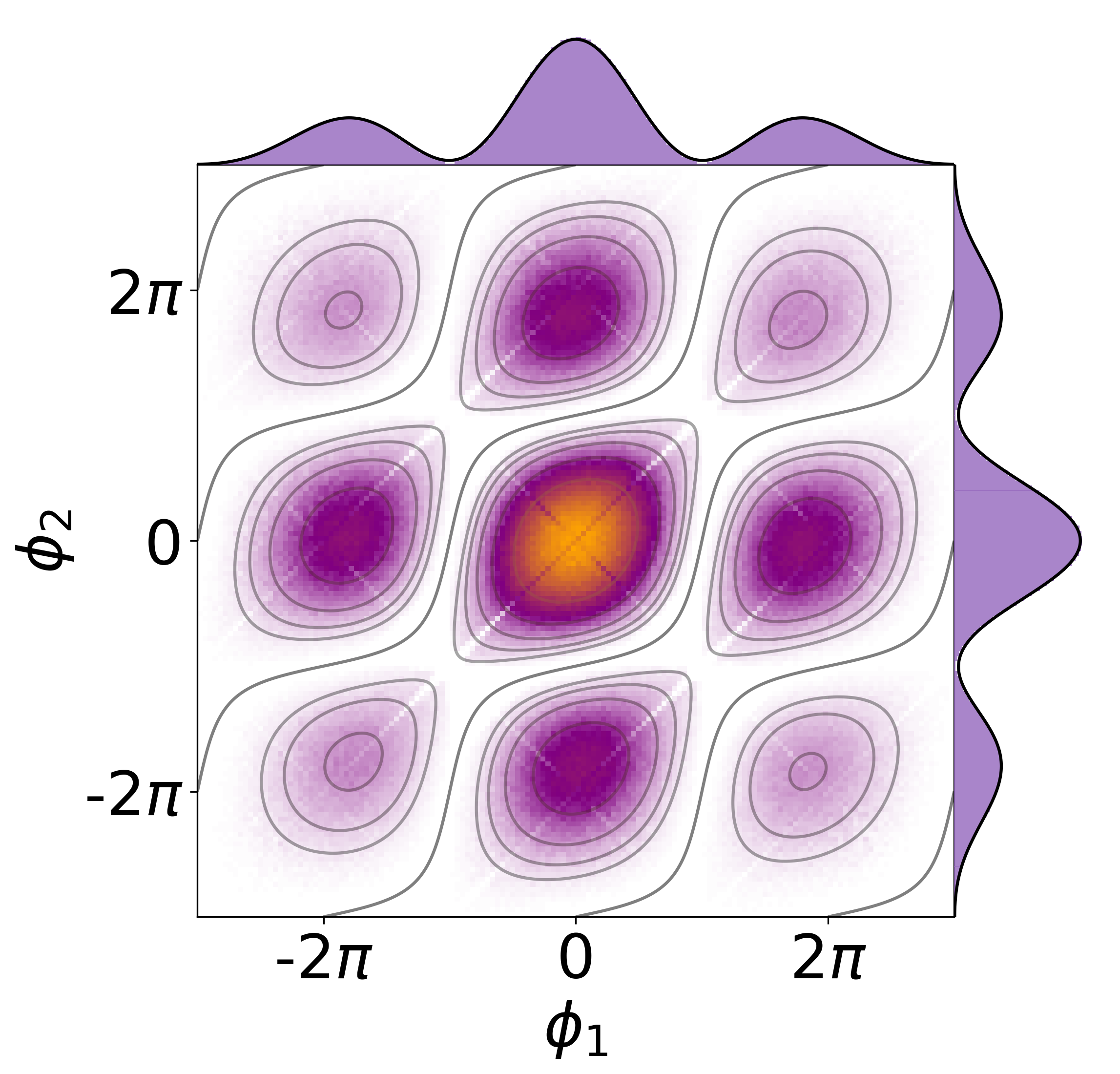}}; 
\node at (3, -0.5) {\includegraphics[width=0.325\linewidth]{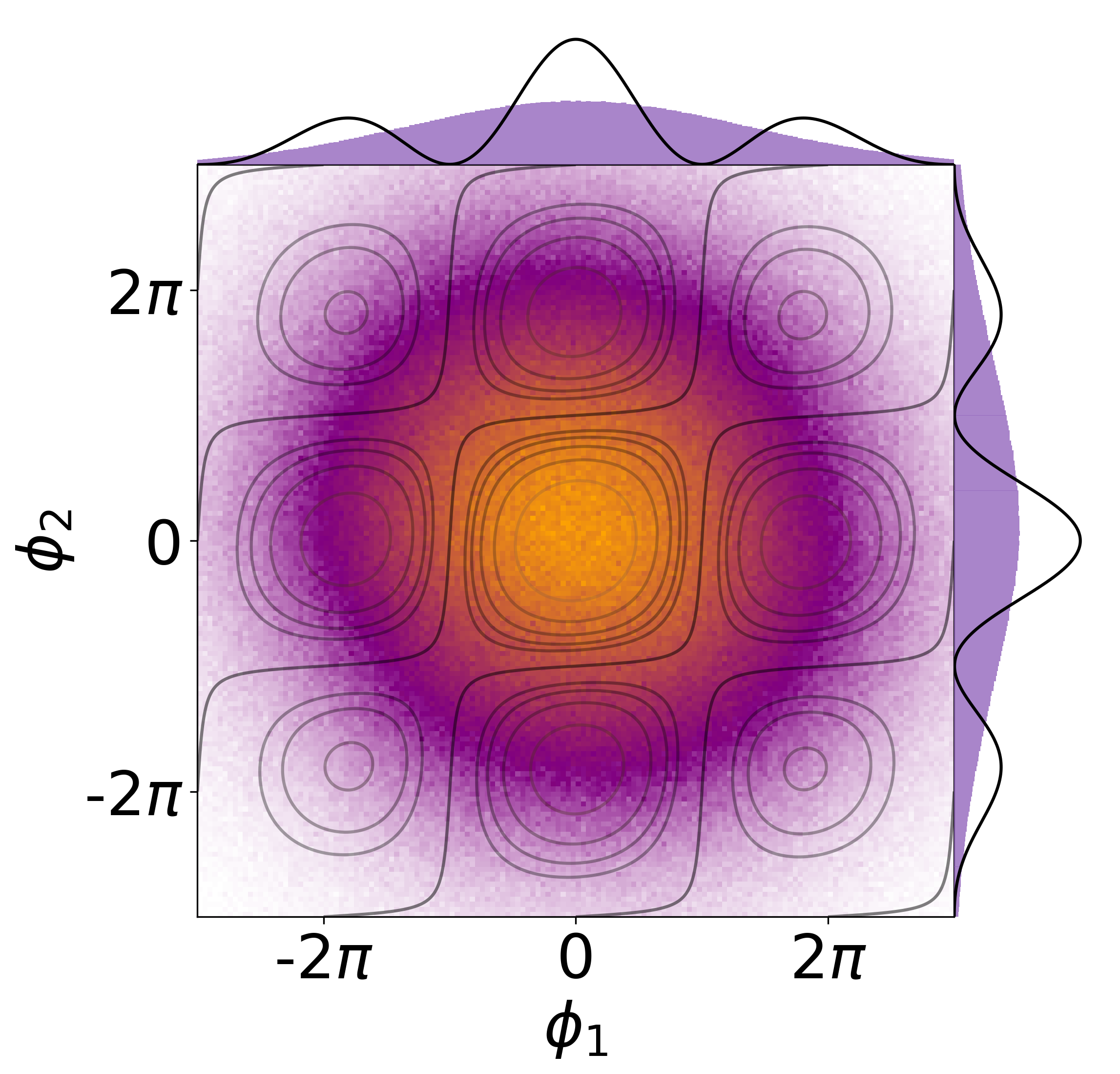}}; 
\node at (8, -0.5) {\includegraphics[width=0.325\linewidth]{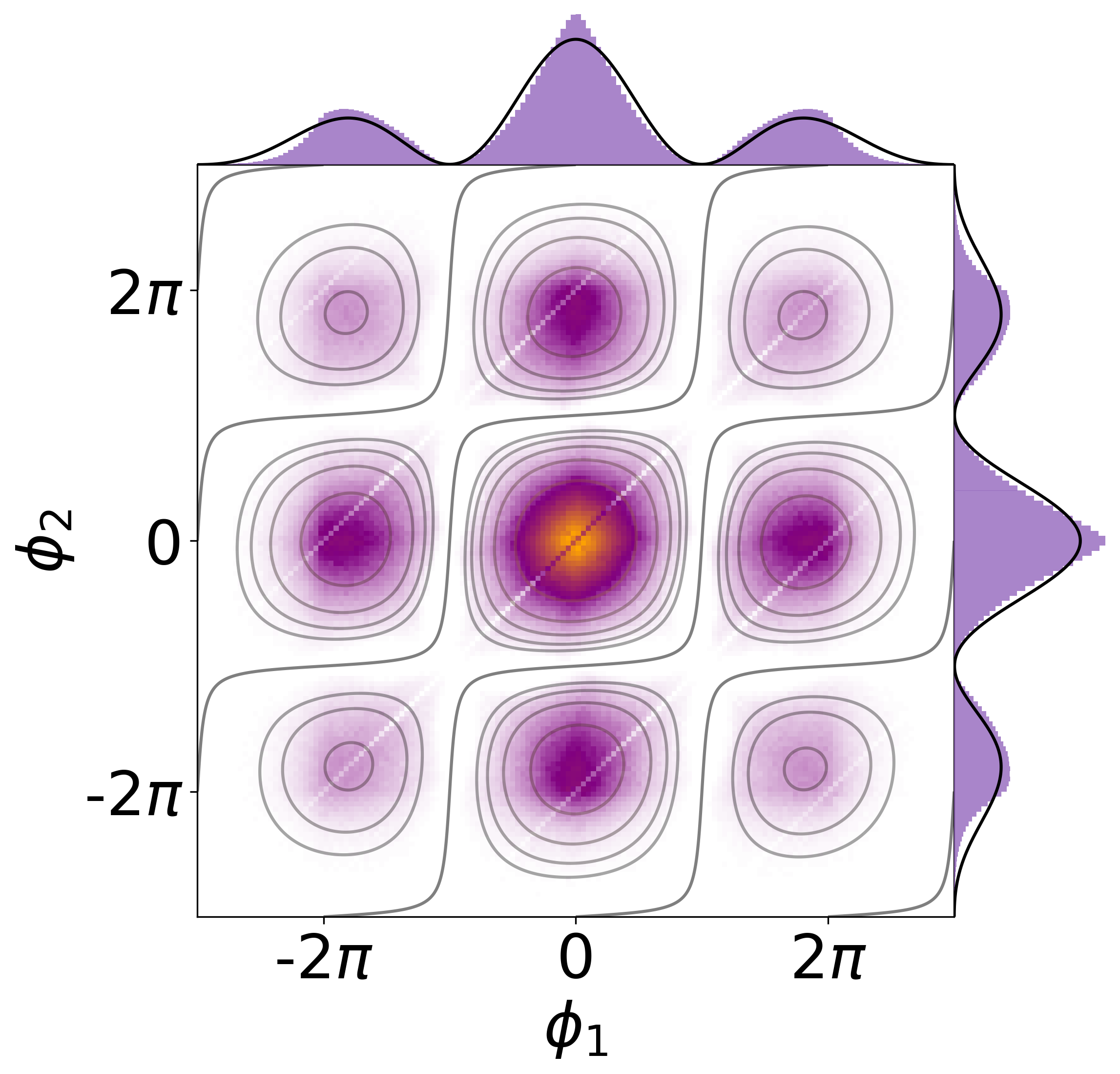}}; 
\node at (13, -0.5) {\includegraphics[width=0.325\linewidth]{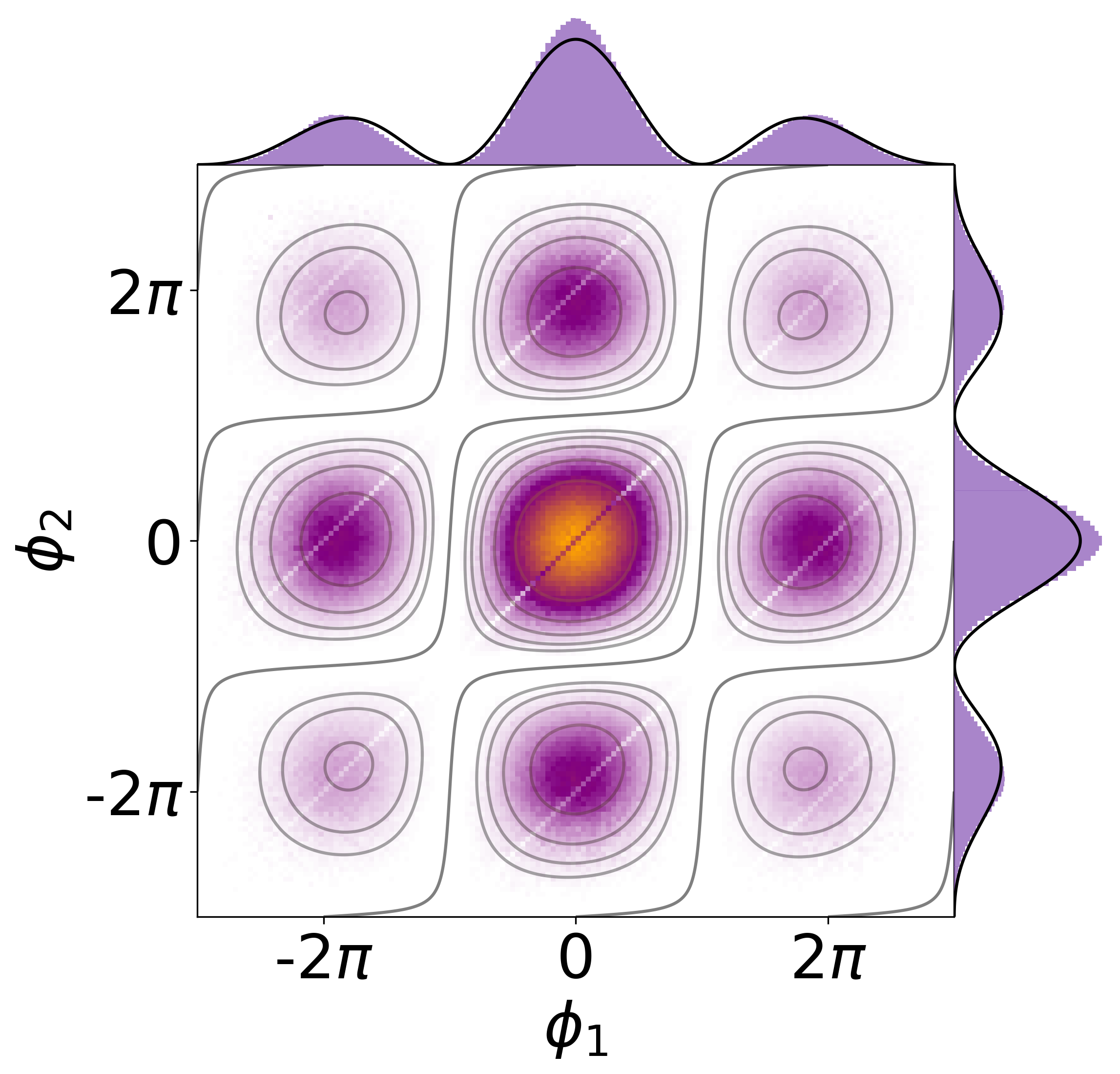}}; 
\node[align=center, anchor=west, rotate=90] at (0, 3.1) {$\mathbf{(N_x,N_t)=(2,1)}$};
\node[align=center, anchor=west, rotate=90] at (0, -1.9) {$\mathbf{(N_x,N_t)=(2,2)}$};
\node[align=center, anchor=west] at (1.15, 7.1) {\textbf{Non-equivariant Flow}};
\node[align=center, anchor=west] at (6.5, 7.1) {\textbf{Equivariant Flow}};
\node[align=center, anchor=west] at (11.8, 7.1) {\textbf{NeuralMCMC}};
\end{tikzpicture}
\caption{Field configurations obtained with non-equivariant (left), equivariant (middle), and equivariant and metropolized (right) normalizing flows. The marginalized magnetization, shown at the top and on the right of each plot, is exact for $N_t=1$ and in the strong-coupling limit for $N_t>1$. The top and bottom rows show results for $(N_x,N_t) = (2,1)$ and $(N_x,N_t) = (2,2)$ lattices, respectively. For $N_t\geq 1$, $\phi_{\{1,2\}}$ is understood to be the sum in the temporal direction, i.e., $\phi_{\{1,2\}} = \sum_{i=1}^{N_t} \phi_{\{1,2\}t}$. While the equivariant approach is able to learn the structure for both lattice sizes to high precision, the non-equivariant flow is unable to learn an approximate to the target distribution for $N_t> 1$. Note that the slightly visible lines in the equivariant distributions originate from a penalty term necessary to keep the normalizing flow bijective while applying the symmetry transformation $T$. The right-most plots on both rows have been obtained by performing metropolization, i.e., filtering i.i.d.\ samples through a metropolis accept-reject step. This ensures that the sampling from the approximate model $q_\theta$ is asymptotically unbiased.}
\label{fig:results}
\end{figure}

\begin{figure}[tb]
    \centering
    \includegraphics[width=0.99\linewidth]{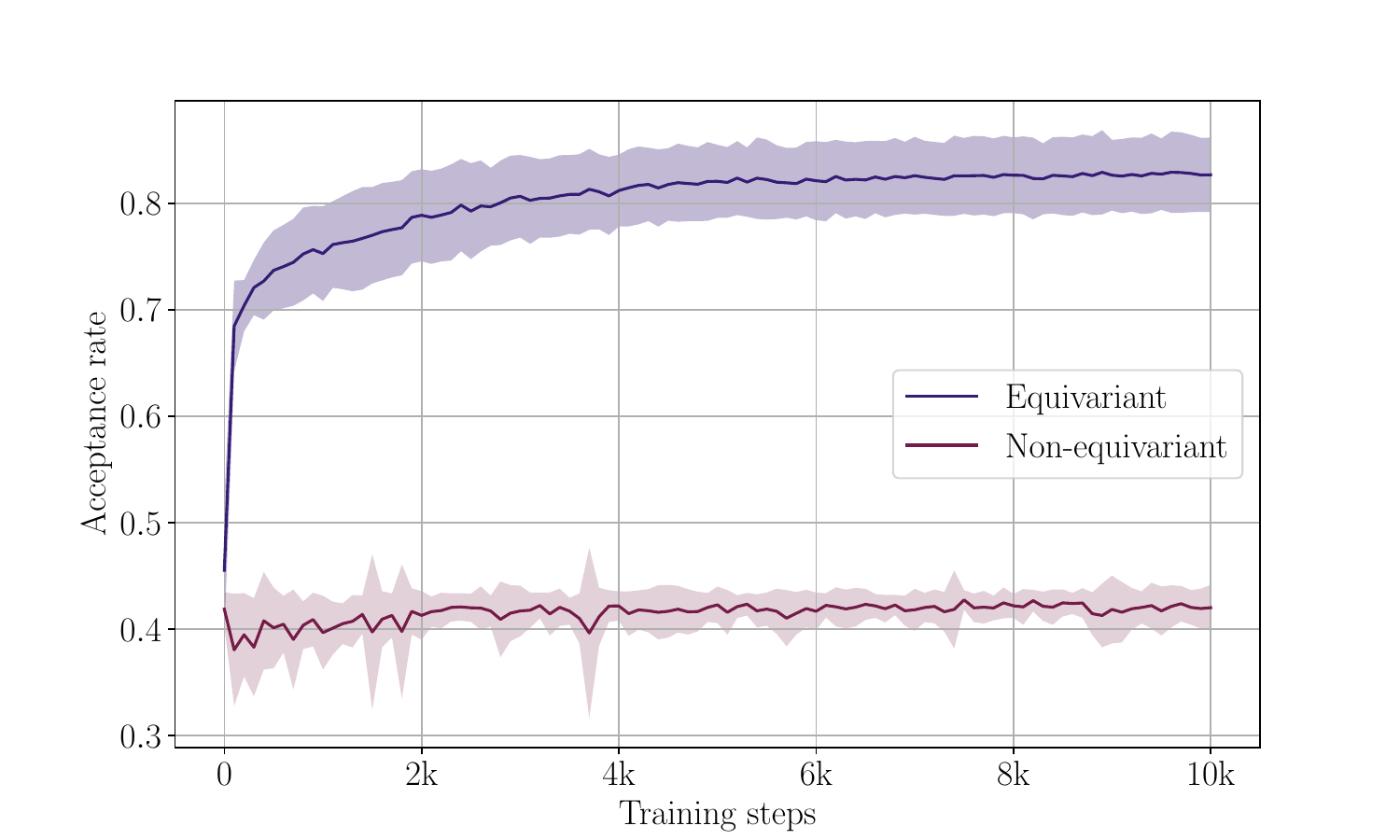}
    \caption{
    Acceptance rate vs.\ training steps for a non-equivariant (red) and an equivariant (purple) normalizing flow for a $(N_x,N_t)=(2,1)$ lattice. The equivariant approach yields acceptance rates above $80\%$ after four thousand training steps, while the non-equivariant only reaches $40\%$. In order to to reach an acceptance rate of$~70\%$, the non-equivariant flow would need more than seven-hundred thousand training steps.}
    \label{fig:acceptance}
\end{figure}

The top row of~\Cref{fig:results} shows our results for a single time-slice, i.e., a $(N_x,N_t)=(2,1)$ system. We see that, in this case, even a \textit{non}-equivariant normalizing flow is able to capture the underlying probability density, as displayed in the left column. However, when making the flow equivariant, i.e., when including the equivariant layer to incorporate the symmetries into the model by design, we observe a notable improvement in both training efficiency as well as the quality of the learned distribution, as shown in the middle column. While the non-equivariant normalizing flow takes 25 hours of training to achieve an acceptance rate of 75\%, the equivariant counterpart converges to an acceptance rate of 85\% already after 16 minutes of training. This significant speedup in training is illustrated in~\Cref{fig:acceptance}. 
Furthermore, when comparing the integrated autocorrelation times in Tab.~\ref{tab:results}, we see a difference of a factor of two between the equivariant and non-equivariant approaches. Remarkably, both approaches show a substantial improvement in the integrated autocorrelation time of roughly two orders of magnitude compared to the standard HMC approach (see the caption of~\Cref{fig:hmc}).

The bottom row of~\Cref{fig:results} shows our results for increasing the temporal extent of the lattice by one, i.e., $N_t=2$. In this case, the non-equivariant normalizing flow is unable to learn the target probability distribution, even after more than 24 hours of training, as shown in the left column. The equivariant normalizing flow, in contrast, is able to approximately learn the target distribution, as shown in the middle column. Here, it achieves an acceptance rate of $69.7 \%$ and an integrated autocorrelation time that is two orders of magnitude smaller than in the standard HMC approach (see~\cref{tab:results} and~\Cref{fig:hmc}), thus indicating a much better sampler. 

As mentioned in~\Cref{sec:symm}, the $\mathbf{Z}_2$- and space-translation symmetries leave the entire action invariant, while the periodicity symmetry is only a symmetry of the fermion determinant, but not of the Gaussian component of the action. In this sense, the periodicity symmetry can be seen as an \textit{approximate} symmetry. The effect of this ``approximation'' can be observed in the magnetization of the equivariant distribution for $(N_x,N_t) = (2,1)$, as shown in the second column of the top row of~\Cref{fig:results}, where the two outer peaks are tilted inwards and, therefore, do not follow the same Gaussian envelope as the analytically obtained contours. However, the third column of the top row of~\Cref{fig:results} demonstrates that this artifact can be removed by using the NeuralMCMC technique.\footnote{Note that the outer peaks for the equivariant distribution of $(N_x,N_t) = (2,2)$ exhibit a similar tilting, see the second column of the bottom row of~\Cref{fig:results}. However, in this case, the analytical solution (solid black line) is not exact for $N_t>1$, thus preventing a similar comparison with the ground truth as for the $(N_x,N_t) = (2,1)$ case.}

\begin{table}[t]
    \centering
    \caption{Integrated autocorrelation times $\tau_\mathrm{int}$ of the equivariant and non-equivariant normalizing flow models for different system sizes of the Hubbard model. Since the non-equivariant approach fails to produce a result for $N_t>1$, a value for $\tau_\mathrm{int}$ cannot be determined in this case.}
    \begin{tabular}{|c|c|c|}
         \hline
         $(N_x, N_t)$ & $\tau_{ \mathrm{int,\,non-equivariant}}$  & $\tau_{ \mathrm{int,\, equivariant}}$ \\
         \hline
         \hline
         (2,1) & $1.52 \pm 0.04$ & $0.71 \pm 0.02$ \\
         (2,2) & - & $1.17\pm0.03$ \\
         \hline
    \end{tabular}
    \label{tab:results}
\end{table}

\section{Conclusion and Outlook}
\label{sec:conclusion}
\noindent
Well-established sampling methods like HMC often suffer from severe ergodicity problems when being applied to the Hubbard model. In this work, we propose to use equivariant normalizing flows to overcome these issues. We present a proof-of-concept demonstration that a normalizing flow can be trained to approximate the target Boltzmann distribution of the Hubbard model and allows to efficiently sample configurations for a $2\times 1$  lattice. Furthermore, we show that incorporating symmetries of the action into an equivariant normalizing flow architecture lowers the training cost, allows to achieve higher acceptance rates, and reduces the integrated autocorrelation times, compared to the non-equivariant approach. 

Nevertheless, making the flow equivariant generally carries the risk of breaking bijectivity, an essential property of a normalizing flow. In order for the normalizing flow to retain this property, one has to restrict the samples to remain \textit{inside} the canonical cell through a penalty term in the loss. While this works well for small systems, the canonical cell is expected to grow with the temporal extent $N_t$, thus reducing the advantage from canonicalizing the prior distribution for a larger number of time-slices. In future work, we will focus on finding a rigorous way to preserve bijectivity of the normalizing flow while incorporating symmetries, in order to use this framework to scale to larger systems. Moreover, we will investigate the performance of normalizing flows for the Hubbard model in different physical regimes, i.e., for various values of the coupling parameters of the theory.

\section*{Acknowledgements}
\noindent
The authors thank Gurtej Kanwar and Daniel Hackett for insightful discussions. This project was supported by the Deutsche Forschungsgemeinschaft (DFG, German Research Foundation) as part of the CRC 1639 NuMeriQS – project no. 511713970.

\bibliographystyle{JHEP}
\bibliography{refs}

\end{document}

%% file: flow.tex
\begin{tikzpicture}[
    node distance=2, very thick,
    flow/.style={shorten >=3, shorten <=3, ->},
    znode/.style={circle, fill=black!10, minimum size=22, inner sep=0},
  ]

  \node[znode] (z0) {$z$};
  \node[znode, right=of z0] (z1) {$y^1$};
  \draw[flow] (z0) -- node[above, midway] {$f^1(z)$} (z1);

  \node[znode, right=2.5 of z1] (zi) {$y^i$};
  \node[znode, right=of zi] (zip1) {$y^{i+1}$};
  \draw[flow] (zi) -- node[above, midway] {$f^{i+1}(y^i)$} (zip1);
  \draw[flow, shorten <=5ex] (z1) -- node[pos=0.16, inner sep=1] {\textbf\dots} node[above, midway] {$f^i(y^{i-1})$} (zi);

  \node[znode, right=2.5 of zip1] (zk) {$y^l$};
  \draw[flow, shorten <=5ex] (zip1) -- node[pos=0.16, inner sep=1] {\textbf\dots} node[above, midway] {$f^l(y^{l-1})$} (zk);
  \node[right=0 of zk, scale=1.2] {$= \phi$};
  \node[outer sep=0, inner sep=0, below=0.2 of z0, label={below:$z \sim q_Z(z)$}] (f0) {\distro{1}{0}{0}};
  \node[outer sep=0, inner sep=0, below=0.2 of zi, label={below:$y^i \sim q^i_\theta(y^{i})$}] (fi) {\distro[70]{1}{1}{0}};
  \node[outer sep=0, inner sep=0, below=0.2 of zk, label={below:$\phi \sim q_\theta(y^{l}) \approx p(\phi) $}] (fk) {\distro[90]{0}{1}{1}};

\end{tikzpicture}

%% file: symmetry_transformation.tex
\begin{tikzpicture}

    \tikzstyle{arrow} = [line width = .4mm,->,>=stealth]
    \tikzstyle{block} = [rectangle, draw, fill=purple!20, text width=8em, text centered, rounded corners, minimum height=4em]

    \node at (-6.4, 0) {\includegraphics[width=4cm]{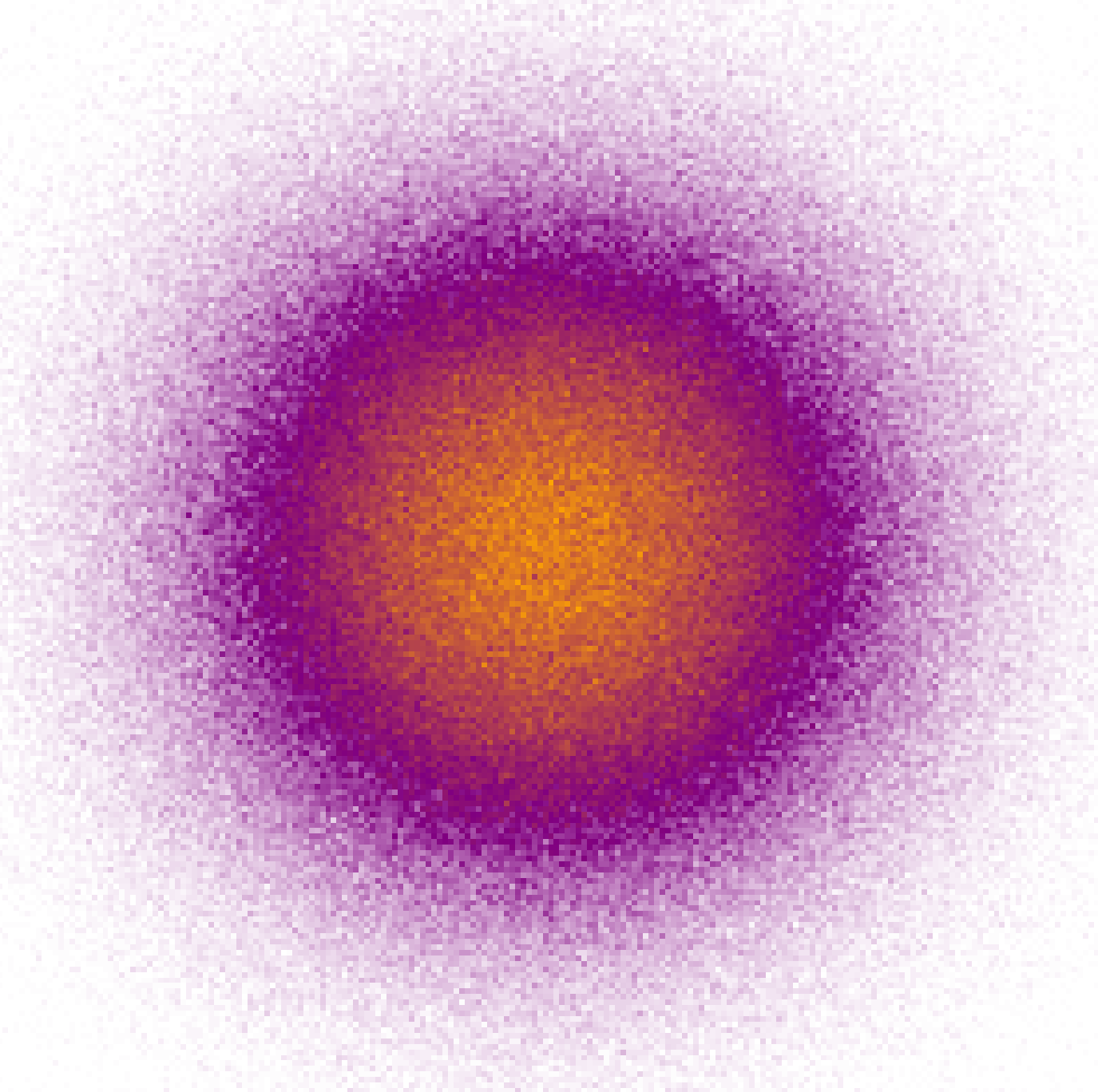}}; 
    \node at (-6.4, -2.75) {Prior Gaussian distribution};


    \node at (0, 2.8) {\includegraphics[height=2.5cm]{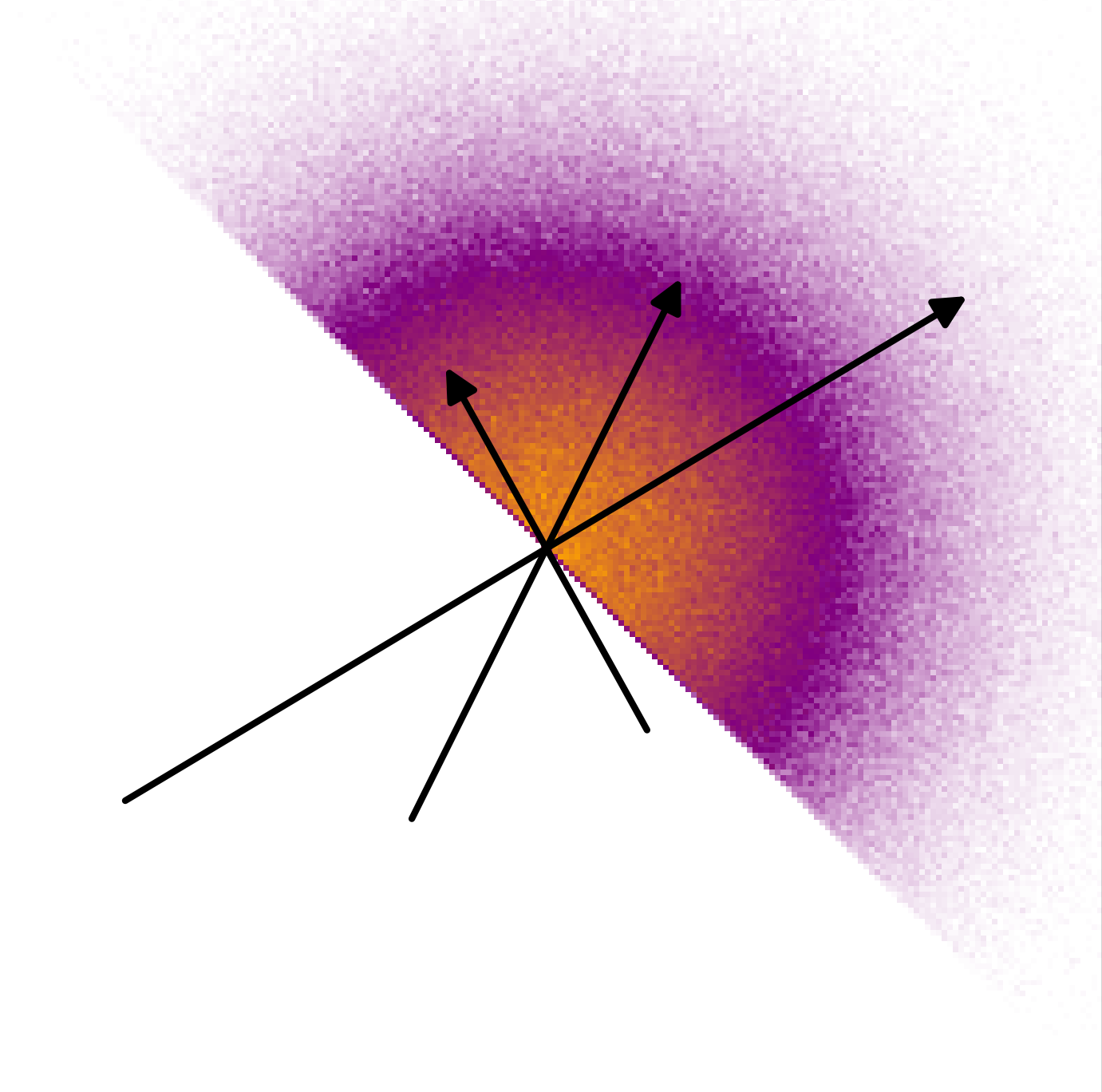}}; 
    \node at (0, 1.5) {$\mathbf{Z}_2$};

    \node at (0, 0) {\includegraphics[height=2.5cm]{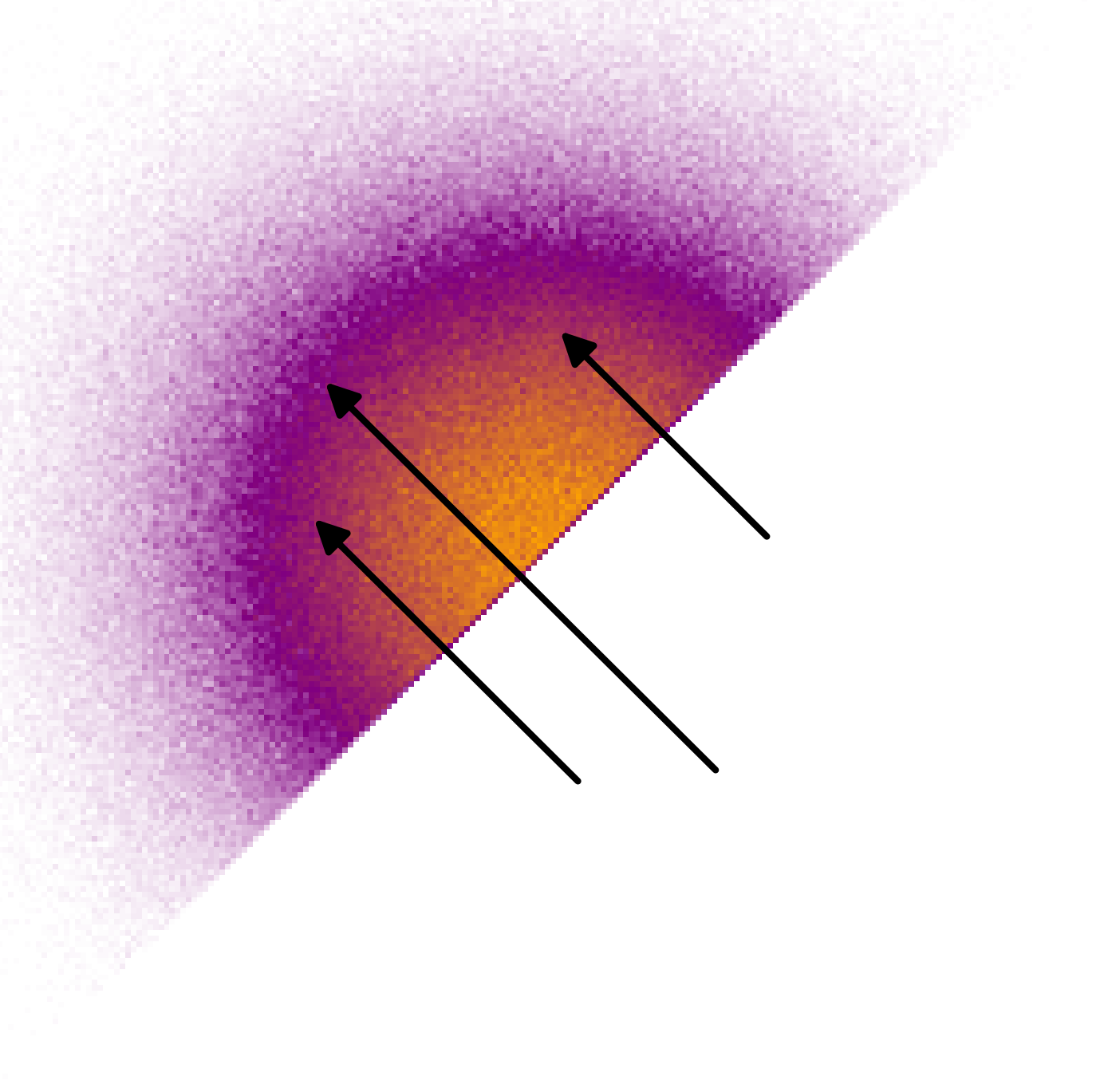}}; 
    \node at (0, -1.4) {Space translation};

    \node at (0, -2.8) {\includegraphics[height=2.5cm]{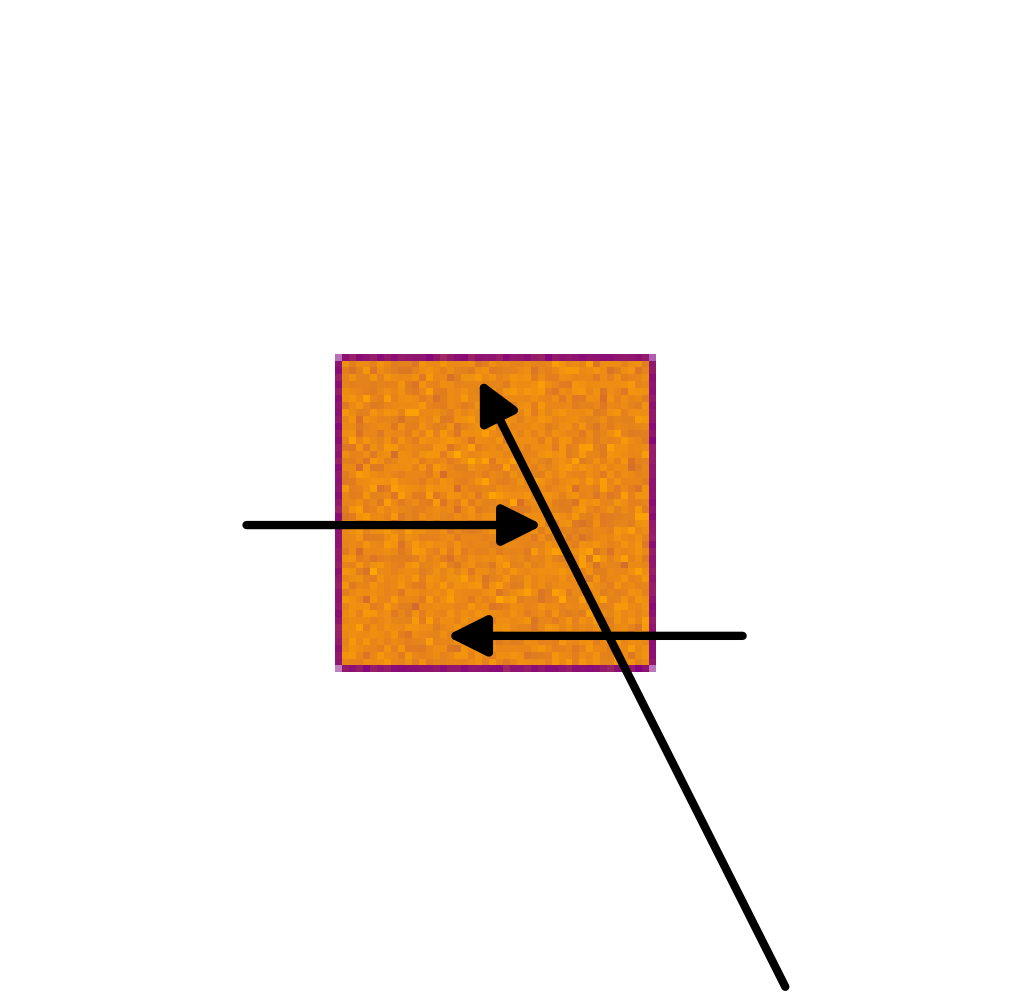}}; 
    \node at (0, -4.2) {Periodicity};

    \node at (6.4, 0) {\includegraphics[width=3cm]{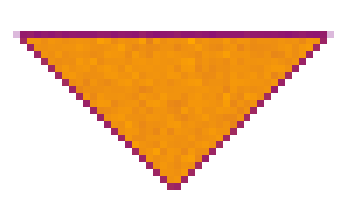}}; 
    \node at (6.4, -2.75) {Canonicalized distribution};

    \draw[arrow] (-4.2, 0) -- (-1.7, 2.8); 
    \draw[arrow] (-4.2, 0) -- (-1.7, 0);   
    \draw[arrow] (-4.2, 0) -- (-1.7, -2.8); 

    \draw[arrow] (-8.1, -1.8) -- (-8.1, 1.5);
    \node at (-8.4, 1.4) {$z_2$};
    \draw[arrow] (-8.1, -1.8) -- (-4.8, -1.8);
    \node at (-4.9, -2.1) {$z_1$};

    \draw[arrow] (1.7, 2.8) -- (4.1567117, 0.04); 
    \draw[arrow] (1.7, 0) -- (4.1567117, 0);   
    \draw[arrow] (1.7, -2.8) -- (4.1567117, -0.04); 

\end{tikzpicture}

%% file: main.bbl
\providecommand{\href}[2]{#2}\begingroup\raggedright\begin{thebibliography}{10}

\bibitem{PhysRevB.100.075141}
J.-L.~Wynen, E.~Berkowitz, C.~K\"orber, T.A.~L\"ahde and T.~Luu, \emph{{Avoiding ergodicity problems in lattice discretizations of the Hubbard model}}, \href{https://doi.org/10.1103/PhysRevB.100.075141}{\emph{Phys. Rev. B} {\bfseries 100} (2019) 075141} [\href{https://arxiv.org/abs/1812.09268}{{\ttfamily 1812.09268}}].

\bibitem{temmen2024overcomingergodicityproblemshybrid}
F.~Temmen, E.~Berkowitz, A.~Kennedy, T.~Luu, J.~Ostmeyer and X.~Yu, \emph{{Overcoming Ergodicity Problems of the Hybrid Monte Carlo Method using Radial Updates}}, {\emph{PoS} {\bfseries LATTICE2024} (2024) 068} [\href{https://arxiv.org/abs/2410.19148}{{\ttfamily 2410.19148}}].

\bibitem{rezende2015variational}
D.~Rezende and S.~Mohamed, \emph{{Variational Inference with Normalizing Flows}},  in \emph{Proceedings of the 32nd International Conference on Machine Learning}, vol.~37, pp.~1530--1538, PMLR, 2015 [\href{https://arxiv.org/abs/1505.05770}{{\ttfamily 1505.05770}}].

\bibitem{kobyzev2020normalizing}
I.~Kobyzev, S.J.~Prince and M.A.~Brubaker, \emph{{Normalizing Flows: An Introduction and Review of Current Methods}}, \href{https://doi.org/10.1109/TPAMI.2020.2992934}{\emph{IEEE Transactions on Pattern Analysis and Machine Intelligence} {\bfseries 43} (2021) 3964} [\href{https://arxiv.org/abs/1908.09257}{{\ttfamily 1908.09257}}].

\bibitem{nfreview}
G.~Papamakarios, E.~Nalisnick, D.J.~Rezende, S.~Mohamed and B.~Lakshminarayanan, \emph{{Normalizing Flows for Probabilistic Modeling and Inference}}, {\emph{J. Mach. Learn. Res.} {\bfseries 22} (2021) } [\href{https://arxiv.org/abs/1912.02762}{{\ttfamily 1912.02762}}].

\bibitem{oord2016pixelrecurrentneuralnetworks}
A.~van~den Oord, N.~Kalchbrenner and K.~Kavukcuoglu, \emph{{Pixel Recurrent Neural Networks}},  in \emph{Proceedings of The 33rd International Conference on Machine Learning}, vol.~48 of \emph{Proceedings of Machine Learning Research}, pp.~1747--1756, PMLR, 20--22 Jun, 2016 [\href{https://arxiv.org/abs/1601.06759}{{\ttfamily 1601.06759}}].

\bibitem{NIPS2016_b1301141}
A.~van~den Oord, N.~Kalchbrenner, L.~Espeholt, K.~Kavukcuoglu, O.~Vinyals and A.~Graves, \emph{{Conditional Image Generation with PixelCNN Decoders}},  in \emph{Advances in Neural Information Processing Systems}, vol.~29, pp.~4797--4805, 2016 [\href{https://arxiv.org/abs/1606.05328}{{\ttfamily 1606.05328}}].

\bibitem{doi:10.1126/science.aaw1147}
F.~Noé, S.~Olsson, J.~Köhler and H.~Wu, \emph{{Boltzmann generators: Sampling equilibrium states of many-body systems with deep learning}}, \href{https://doi.org/10.1126/science.aaw1147}{\emph{Science} {\bfseries 365} (2019) eaaw1147} [\href{https://arxiv.org/abs/1812.01729}{{\ttfamily 1812.01729}}].

\bibitem{PhysRevD.100.034515}
M.S.~Albergo, G.~Kanwar and P.E.~Shanahan, \emph{{Flow-based generative models for Markov chain Monte Carlo in lattice field theory}}, \href{https://doi.org/10.1103/PhysRevD.100.034515}{\emph{Phys. Rev. D} {\bfseries 100} (2019) 034515} [\href{https://arxiv.org/abs/1904.12072}{{\ttfamily 1904.12072}}].

\bibitem{PhysRevLett.126.032001}
K.A.~Nicoli, C.J.~Anders, L.~Funcke, T.~Hartung, K.~Jansen, P.~Kessel et~al., \emph{{Estimation of Thermodynamic Observables in Lattice Field Theories with Deep Generative Models}}, \href{https://doi.org/10.1103/PhysRevLett.126.032001}{\emph{Phys. Rev. Lett.} {\bfseries 126} (2021) 032001} [\href{https://arxiv.org/abs/2007.07115}{{\ttfamily 2007.07115}}].

\bibitem{Caselle:2022acb}
M.~Caselle, E.~Cellini, A.~Nada and M.~Panero, \emph{{Stochastic normalizing flows as non-equilibrium transformations}}, \href{https://doi.org/10.1007/JHEP07(2022)015}{\emph{JHEP} {\bfseries 07} (2022) 015} [\href{https://arxiv.org/abs/2201.08862}{{\ttfamily 2201.08862}}].

\bibitem{cranmer2023advances}
K.~Cranmer, G.~Kanwar, S.~Racani{\`e}re, D.J.~Rezende and P.E.~Shanahan, \emph{{Advances in machine-learning-based sampling motivated by lattice quantum chromodynamics}}, \href{https://doi.org/10.1038/s42254-023-00616-w}{\emph{Nature Reviews Physics} {\bfseries 5} (2023) 526} [\href{https://arxiv.org/abs/2309.01156}{{\ttfamily 2309.01156}}].

\bibitem{PhysRevLett.122.080602}
D.~Wu, L.~Wang and P.~Zhang, \emph{{Solving Statistical Mechanics Using Variational Autoregressive Networks}}, \href{https://doi.org/10.1103/PhysRevLett.122.080602}{\emph{Phys. Rev. Lett.} {\bfseries 122} (2019) 080602} [\href{https://arxiv.org/abs/1809.10606}{{\ttfamily 1809.10606}}].

\bibitem{PhysRevE.101.023304}
K.A.~Nicoli, S.~Nakajima, N.~Strodthoff, W.~Samek, K.-R.~M\"uller and P.~Kessel, \emph{{Asymptotically unbiased estimation of physical observables with neural samplers}}, \href{https://doi.org/10.1103/PhysRevE.101.023304}{\emph{Phys. Rev. E} {\bfseries 101} (2020) 023304} [\href{https://arxiv.org/abs/1910.13496}{{\ttfamily 1910.13496}}].

\bibitem{Caselle:2023uel}
M.~Caselle, E.~Cellini and A.~Nada, \emph{{Sampling Nambu-Goto theory using Normalizing Flows}}, \href{https://doi.org/10.22323/1.453.0015}{\emph{PoS} {\bfseries LATTICE2023} (2024) 015} [\href{https://arxiv.org/abs/2309.14983}{{\ttfamily 2309.14983}}].

\bibitem{Caselle:2023mvh}
M.~Caselle, E.~Cellini and A.~Nada, \emph{{Sampling the lattice Nambu-Goto string using Continuous Normalizing Flows}}, \href{https://doi.org/10.1007/JHEP02(2024)048}{\emph{JHEP} {\bfseries 02} (2024) 048} [\href{https://arxiv.org/abs/2307.01107}{{\ttfamily 2307.01107}}].

\bibitem{gebauer1}
N.~Gebauer, M.~Gastegger and K.~Sch\"{u}tt, \emph{{Symmetry-adapted generation of 3d point sets for the targeted discovery of molecules}},  in \emph{Advances in Neural Information Processing Systems}, vol.~32, 2019 [\href{https://arxiv.org/abs/1906.00957}{{\ttfamily 1906.00957}}].

\bibitem{gebauer3}
N.W.A.~Gebauer, M.~Gastegger, S.S.P.~Hessmann, K.-R.~M{\"u}ller and K.T.~Sch{\"u}tt, \emph{{Inverse design of 3d molecular structures with conditional generative neural networks}}, \href{https://doi.org/10.1038/s41467-022-28526-y}{\emph{Nature Communications} {\bfseries 13} (2022) 973}.

\bibitem{bialas2024r}
P.~Bia\l{}as, P.~Korcyl, T.~Stebel and D.~Zapolski, \emph{{R\'enyi entanglement entropy of a spin chain with generative neural networks}}, \href{https://doi.org/10.1103/PhysRevE.110.044116}{\emph{Phys. Rev. E} {\bfseries 110} (2024) 044116} [\href{https://arxiv.org/abs/2406.06193}{{\ttfamily 2406.06193}}].

\bibitem{Bulgarelli:2024yrz}
A.~Bulgarelli, E.~Cellini, K.~Jansen, S.~K\"uhn, A.~Nada, S.~Nakajima et~al., \emph{{Flow-based Sampling for Entanglement Entropy and the Machine Learning of Defects}}, {\emph{ArXiv e-prints} (2024) } [\href{https://arxiv.org/abs/2410.14466}{{\ttfamily 2410.14466}}].

\bibitem{Nicoli:2021inv}
K.A.~Nicoli, C.J.~Anders, L.~Funcke, T.~Hartung, K.~Jansen, P.~Kessel et~al., \emph{{Machine Learning of Thermodynamic Observables in the Presence of Mode Collapse}}, \href{https://doi.org/10.22323/1.396.0338}{\emph{PoS} {\bfseries LATTICE2021} (2022) 338} [\href{https://arxiv.org/abs/2111.11303}{{\ttfamily 2111.11303}}].

\bibitem{PhysRevD.108.114501}
K.A.~Nicoli, C.J.~Anders, T.~Hartung, K.~Jansen, P.~Kessel and S.~Nakajima, \emph{{Detecting and mitigating mode-collapse for flow-based sampling of lattice field theories}}, \href{https://doi.org/10.1103/PhysRevD.108.114501}{\emph{Phys. Rev. D} {\bfseries 108} (2023) 114501} [\href{https://arxiv.org/abs/2302.14082}{{\ttfamily 2302.14082}}].

\bibitem{PhysRevLett.125.121601}
G.~Kanwar, M.S.~Albergo, D.~Boyda, K.~Cranmer, D.C.~Hackett, S.~Racani\`ere et~al., \emph{{Equivariant Flow-Based Sampling for Lattice Gauge Theory}}, \href{https://doi.org/10.1103/PhysRevLett.125.121601}{\emph{Phys. Rev. Lett.} {\bfseries 125} (2020) 121601} [\href{https://arxiv.org/abs/2003.06413}{{\ttfamily 2003.06413}}].

\bibitem{kohler2020equivariant}
J.~K{\"o}hler, L.~Klein and F.~No{\'e}, \emph{{Equivariant flows: exact likelihood generative learning for symmetric densities}},  in \emph{International conference on machine learning}, pp.~5361--5370, PMLR, 2020 [\href{https://arxiv.org/abs/2006.02425}{{\ttfamily 2006.02425}}].

\bibitem{hubbard}
J.~Hubbard, \emph{{Electron correlations in narrow energy bands}}, \href{https://doi.org/10.1098/rspa.1963.0204}{\emph{Proceedings of the Royal Society of London. Series A. Mathematical and Physical Sciences} {\bfseries 276} (1963) 238}.

\bibitem{Arovas_2022}
D.P.~Arovas, E.~Berg, S.A.~Kivelson and S.~Raghu, \emph{{The Hubbard Model}}, \href{https://doi.org/10.1146/annurev-conmatphys-031620-102024}{\emph{Annual Review of Condensed Matter Physics} {\bfseries 13} (2022) 239} [\href{https://arxiv.org/abs/2103.12097}{{\ttfamily 2103.12097}}].

\bibitem{trotter1959product}
H.F.~Trotter, \emph{{On the product of semi-groups of operators}}, \href{https://doi.org/10.2307/2033649}{\emph{Proceedings of the American Mathematical Society} {\bfseries 10} (1959) 545}.

\bibitem{10.1143/PTP.56.1454}
M.~Suzuki, \emph{{Relationship between d-Dimensional Quantal Spin Systems and (d+1)-Dimensional Ising Systems: Equivalence, Critical Exponents and Systematic Approximants of the Partition Function and Spin Correlations}}, \href{https://doi.org/10.1143/PTP.56.1454}{\emph{Progress of Theoretical Physics} {\bfseries 56} (1976) 1454}.

\bibitem{Hubbard:1959ub}
J.~Hubbard, \emph{{Calculation of Partition Functions}}, \href{https://doi.org/10.1103/PhysRevLett.3.77}{\emph{Phys. Rev. Lett.} {\bfseries 3} (1959) 77}.

\bibitem{brower2012hybridmontecarlosimulation}
R.C.~Brower, D.~Schaich and C.~Rebbi, \emph{{Hybrid Monte Carlo simulation on the graphene hexagonal lattice}}, \href{https://doi.org/10.22323/1.139.0056}{\emph{PoS} {\bfseries Lattice 2011} (2012) 056} [\href{https://arxiv.org/abs/1204.5424}{{\ttfamily 1204.5424}}].

\bibitem{Ulybyshev_2013}
M.V.~Ulybyshev, P.V.~Buividovich, M.I.~Katsnelson and M.I.~Polikarpov, \emph{{Monte Carlo Study of the Semimetal-Insulator Phase Transition in Monolayer Graphene with a Realistic Interelectron Interaction Potential}}, \href{https://doi.org/10.1103/PhysRevLett.111.056801}{\emph{Phys. Rev. Lett.} {\bfseries 111} (2013) 056801} [\href{https://arxiv.org/abs/1304.3660}{{\ttfamily 1304.3660}}].

\bibitem{Smith_2014}
D.~Smith and L.~von Smekal, \emph{{Monte Carlo simulation of the tight-binding model of graphene with partially screened Coulomb interactions}}, \href{https://doi.org/10.1103/PhysRevB.89.195429}{\emph{Phys. Rev. B} {\bfseries 89} (2014) 195429} [\href{https://arxiv.org/abs/1403.3620}{{\ttfamily 1403.3620}}].

\bibitem{Luu_2016}
T.~Luu and T.A.~L\"ahde, \emph{{Quantum Monte Carlo calculations for carbon nanotubes}}, \href{https://doi.org/10.1103/PhysRevB.93.155106}{\emph{Phys. Rev. B} {\bfseries 93} (2016) 155106} [\href{https://arxiv.org/abs/1511.04918}{{\ttfamily 1511.04918}}].

\bibitem{Duane:1987de}
S.~Duane, A.D.~Kennedy, B.J.~Pendleton and D.~Roweth, \emph{{Hybrid Monte Carlo}}, \href{https://doi.org/10.1016/0370-2693(87)91197-X}{\emph{Phys. Lett. B} {\bfseries 195} (1987) 216}.

\bibitem{dinh2014nice}
L.~Dinh, D.~Krueger and Y.~Bengio, \emph{{NICE:} non-linear independent components estimation},  in \emph{3rd International Conference on Learning Representations (Workshop Track Proceedings)}, 2015 [\href{https://arxiv.org/abs/1410.8516}{{\ttfamily 1410.8516}}].

\bibitem{dinh2017densityestimationusingreal}
L.~Dinh, J.~Sohl-Dickstein and S.~Bengio, \emph{{Density estimation using Real {NVP}}},  in \emph{International Conference on Learning Representations}, 2017 [\href{https://arxiv.org/abs/1605.08803}{{\ttfamily 1605.08803}}].

\bibitem{10.1214/aoms/1177729694}
S.~Kullback and R.A.~Leibler, \emph{{On Information and Sufficiency}}, \href{https://doi.org/10.1214/aoms/1177729694}{\emph{The Annals of Mathematical Statistics} {\bfseries 22} (1951) 79}.

\bibitem{boyda2021sampling}
D.~Boyda, G.~Kanwar, S.~Racani\`ere, D.J.~Rezende, M.S.~Albergo, K.~Cranmer et~al., \emph{{Sampling using $\mathrm{SU}(N)$ gauge equivariant flows}}, \href{https://doi.org/10.1103/PhysRevD.103.074504}{\emph{Phys. Rev. D} {\bfseries 103} (2021) 074504} [\href{https://arxiv.org/abs/2008.05456}{{\ttfamily 2008.05456}}].

\bibitem{Nicoli:2023rcd}
K.A.~Nicoli, C.J.~Anders, L.~Funcke, K.~Jansen, S.~Nakajima and P.~Kessel, \emph{{NeuLat: a toolbox for neural samplers in lattice field theories}}, \href{https://doi.org/10.22323/1.453.0286}{\emph{PoS} {\bfseries LATTICE2023} (2024) 286}.

\end{thebibliography}\endgroup
